\documentclass[10pt]{article}
\usepackage{amsfonts}
\usepackage{epsfig,amsmath,graphicx,amssymb,overpic,setspace}
\usepackage{mathrsfs}
\usepackage{epsfig,amsmath,graphicx,amssymb,overpic,url,cite}
\usepackage{epstopdf}
\usepackage{colordvi}
\usepackage{color}

\def\be{\begin{equation}}
\def\ee{\end{equation}}
\def\bee{\begin{eqnarray}}
\def\ene{\end{eqnarray}}
\def\bes{\begin{subequations}}
\def\ees{\end{subequations}}

\newcommand{\PT}{ \mathcal{PT}}
\def\be{\begin{equation}}
\def\ee{\end{equation}}
\def\bee{\begin{eqnarray}}
\def\ene{\end{eqnarray}}
\def\bes{\begin{subequations}}
\def\ees{\end{subequations}}

\begin{document}

\baselineskip=8pt \renewcommand {\thefootnote}{\dag}
\renewcommand
{\thefootnote}{\ddag} \renewcommand {\thefootnote}{ }

\pagestyle{plain}

\baselineskip=12pt %\leftline{} \vspace{-.3in} {Stable b
\noindent{\large \textbf{Families of stable solitons and excitations in the $%
\mathcal{PT}$-symmetric nonlinear Schr\"odinger equations with
position-dependent effective masses}} \newline

%\begin{center}

\noindent \textbf{Yong Chen$^{1,2}$, Zhenya Yan$^{1,2}$}$^{\dag }$\footnote{$%
^{\dag }$ Correspondence and requests for materials should be addressed to
Z.Y. (zyyan@mmrc.iss.ac.cn)}, \textbf{Dumitru Mihalache$^{3}$} \& \textbf{%
Boris A. Malomed$^{4,5}$} \\[0.1in]
{\small $^{1}$Key Laboratory of Mathematics Mechanization, Institute of
Systems Science, AMSS, Chinese Academy of Sciences, Beijing 100190, China
\newline
$^{2}$School of Mathematical Sciences, University of Chinese Academy of
Sciences, Beijing 100049, China \newline
$^{3}$Department of Theoretical Physics, Horia Hulubei National Institute of
Physics and Nuclear Engineering, PO Box MG-6, Bucharest, Romania \newline
$^{4}$Department of Physical Electronics, School of Electrical Engineering,
Tel Aviv University, Tel Aviv 59978, Israel \newline
$^{5}$Laboratory of Nonlinear-Optical Informatics, ITMO University, St.
Petersburg 197101, Russia}

%\end{center}

\vspace{0.15in}

{\baselineskip=15pt }

%\begin{abstract}  \baselineskip=14pt

\noindent \textbf{Since the parity-time-($\mathcal{PT}$-) symmetric quantum
mechanics was put forward, fundamental properties of some linear and
nonlinear models with $\mathcal{PT}$-symmetric potentials have been investigated. However, previous studies
of $\mathcal{PT}$-symmetric waves were limited to constant diffraction
coefficients in the ambient medium. Here we address effects of variable
diffraction coefficient on the beam dynamics in
nonlinear media with generalized $\mathcal{PT}$-symmetric Scarf-II potentials. 
The
broken linear $\mathcal{PT}$ symmetry phase may enjoy a restoration with the
growing diffraction parameter. Continuous families of one- and
two-dimensional solitons are found to be stable. Particularly, some stable solitons are analytically found. 
The existence
range and propagation dynamics of the solitons are identified.
Transformation of the solitons by means of adiabatically varying parameters,
and collisions between solitons are studied too. We also explore the
evolution of constant-intensity waves in a model combining the variable
diffraction coefficient and complex potentials with globally balanced gain
and loss, which are more general than $\mathcal{PT}$-symmetric ones, but
feature similar properties. Our results may suggest new experiments for }$%
\mathcal{PT}$\textbf{-symmetric nonlinear waves in nonlinear nonuniform
optical media.}

\vspace{0.3in}

The Hamiltonians in the quantum mechanics are usually required to be
Hermitian, which secures the corresponding spectra to be real~\cite{qm}.
Nevertheless, it had been demonstrated by Bender and Boettcher in 1998 that
non-Hermitian Hamiltonians obeying the parity-time ($\mathcal{PT}$) symmetry
may also produce entirely real spectra~\cite{bender1, dorey, bender2,
bender3, bender4, review,ptqm}. The $\mathcal{PT}$ symmetry implies that the
real and imaginary parts of the complex-valued potential, $U(x)=V(x)+iW(x)$,
are, respectively, even and odd functions of the coordinate: $V(x)=V(-x)$, $%
W(-x)=-W(x)$~\cite{bender1}. For a given real part of the potential, the
spectrum of most $\mathcal{PT}$-symmetric systems remains real, as long as
the amplitude of the imaginary component of the potential is kept below a
certain critical value (the $\mathcal{PT}$-symmetry-breaking threshold);
nevertheless, dynamical models featuring unbreakable $\mathcal{PT}$ symmetry
are known too \cite{unbreakable}. Pioneering theoretical works had predicted
a possibility to realize the $\mathcal{PT}$-symmetric wave propagation in
optical media with symmetrically placed gain and loss elements \cite%
{theo1,theo2,theo3,theo4,theo5}, which was followed by the experimental
implementation in optical and atomic settings, including synthetic photonic
lattices, metamaterials, microring lasers, whispering-gallery microcavities,
and optically induced atomic lattices \cite{exp1,exp2,exp3,exp4,exp5,exp6,
exp7,exp8}. In particular, phase transitions between regions of the unbroken
and broken $\mathcal{PT}$ symmetry have been observed in many experiments.

On the theoretical side, the consideration of $\mathcal{PT}$-symmetric
potentials in both one- and multi-dimensional linear and nonlinear Schr\"{o}%
dinger (NLS) or Gross-Pitaevskii (GP) equations has revealed many remarkable
$\mathcal{PT}$-symmetry-breaking phenomena, including several models that
give rise to $\mathcal{PT}$-symmetric solitons \cite{t1, shi2011, t2,
Radik,t3, t4, Barash,t5, t6, t7, t8, t8a,t9, Raymond, Jennie, t10, t11, t12,
t13, t13a, t13b, t14, t15,t16,t17,t17a,t17b,t18, t19,t20}. It is commonly known that the
soliton theory has been widely applied to fluid mechanics, plasma physics,
Bose-Einstein condensates (BECs), nonlinear optics, and many other fields.
In particular, optical solitons can utilize the nonlinearity in optical
fibers to balance the group-velocity dispersion, thus stably propagating in
long-scale telecommunication links. More recently, stable $\mathcal{PT}$%
-symmetric solitons were also investigated in the third-order NLS equation~%
\cite{yansr16}, the generalized GP equation with a variable group-velocity
coefficient~\cite{yanchaos16}, and the derivative NLS equation~\cite%
{yanstable2017}. The vast work performed in the field of nonlinear waves in $%
\mathcal{PT}$-symmetric systems has been summarized in two recent
comprehensive reviews \cite{pt-review-1,pt-review}.

As mentioned above, the usual one-dimensional (1D) $\mathcal{PT}$-symmetric
Hamiltonian is $\mathcal{H}=-\partial _{x}^{2}+V(x)+iW(x)$, with $V(-x)=V(x)$
and $W(-x)=-W(x)$~\cite{bender1}. However, physics of semiconductors gives
rise to a Hamiltonian in which the effective mass of collective excitations
\cite{em1}, $M(x)$, may be variable (position-dependent) \cite%
{em2,em3,em4,em5,em45,em6,em7,em8}: $\mathcal{H}_{M}=-\frac{1}{2}\hbar
^{2}\partial _{x}\left( \frac{1}{M(x)}\partial _{x}\right) +V(x)$.

To the best of our knowledge, Hamiltonians combining a variable effective
mass and complex-valued $\mathcal{PT}$-symmetric potentials, in particular, $%
\mathcal{H}_{M}^{\left( \mathcal{PT}\right) }=-\frac{1}{2}\hbar ^{2}\partial
_{x}\left( \frac{1}{M(x)}\partial _{x}\right) +V(x)+iW(x)$, have not been
studied yet. In fact, this model represents more general settings than the
above-mentioned one occurring in semiconductors. Indeed, the effective mass
for collective excitations propagating in lattice media is determined by
local properties of the underlying lattice \cite{Inguscio,Pitaevskii}, which
may be nonuniform in many situations, thus making the effective mass
position-dependent. In the context of optics, essentially the same
Hamiltonian governs the light propagation in the spatial domain, where the
effective diffraction coefficient, $1/M$, can be also made $x$-dependent in
nonuniform photonic lattices \cite{Silberberg,Longhi}.

The goal of the present work is to introduce 1D and 2D NLS models with such
Hamiltonians, incorporating a particular physically relevant $\mathcal{PT}$%
-symmetric potential, namely, a generalized Scarf-II potentials (i.e., a
hyperbolic version of the quantum-mechanical potential introduced by Scarf
\cite{Scarf}). We reveal characteristic properties of both 1D and 2D linear
and nonlinear modes in such models, including solitons in the case of cubic
nonlinearity, which are quite generic and can be extended to other
physically relevant complex-valued $\mathcal{PT}$-symmetric potentials.

\vspace{0.1in} \noindent{\large \textbf{Results}}

\noindent \textbf{1D $\mathcal{PT}$-symmetric nonlinear waves in the
effective diffraction.} In Kerr-type nonlinear media with the complex-valued
$\mathcal{PT}$-symmetric potential and the effective diffraction coefficient
defined by the position-dependent mass, $m(x)\equiv 1/M(x)$ (in particular,
it represents the above-mentioned variable diffraction coefficient in
optics), the scaled 1D modified NLS equation for the wave function $\psi
(x,z)$ is%
\begin{equation}
i\frac{\partial \psi }{\partial z}=\left[ -\frac{\partial }{\partial x}%
\left( m(x)\frac{\partial }{\partial x}\right) +V(x)+iW(x)-g|\psi |^{2}%
\right] \psi ,  \label{nls}
\end{equation}%
where $z$ is the propagation distance, $x$ is the transverse coordinate, and
$g$ is the strength of the Kerr nonlinearity. Replacing $z$ by time $t$,
Eq.~(\ref{nls}) may be used as the GP equation for the BEC\ wave function,
with the effective mass affected by a nonuniform optical lattice \cite{BEC}.
In the spatial-domain optics, real
potential $V(x)$ represents the local modification of the refractive index,
whereas $W(x)$ stands for the transverse symmetric distribution of the
optical gain ($W>0$) and loss ($W<0$). Under the conditions that $V(x)+iW(x)$
is a $\mathcal{PT}$-symmetric potential and $m(x)$ is an even function of $x$%
, it is easy to see that Eq.~(\ref{nls}) is invariant under the action of $%
\mathcal{PT}$-symmetric transformation, where the spatial-reflection
operator $\mathcal{P}$ and time-reversal operator $\mathcal{T}$ are defined
as usual \cite{bender1}, $\mathcal{P}:\,x\rightarrow -x;\,\mathcal{T}%
:i\rightarrow -i,\,z\rightarrow -z$ . Equation~(\ref{nls}) may be rewritten
in the variational form, $i\psi _{z}=\delta \mathcal{H}(\psi )/\delta \psi
^{\ast }$, with a non-Hermitian but $\mathcal{PT}$-symmetric field
Hamiltonian, $\mathcal{H}(\psi )=\int_{-\infty }^{+\infty }\{m(x)|\psi
_{x}|^{2}+[V(x)+iW(x)]|\psi |^{2}-\frac{g}{2}|\psi |^{4}\}dx$, where the
asterisk denotes the complex conjugate. Further, the power (norm) of the
wave function, $P(z)=\int_{-\infty }^{+\infty }|\psi (x,z)|^{2}dx$, evolves
according to equation $dP/dz=2\int_{-\infty }^{+\infty }W(x)|\psi |^{2}dx$,
which can be immediately deduced from Eq. (\ref{nls}).

%%%%%%%%%%%%%%%%%%%%%%%%%%%%%%%%%%%%%%%%%%%%%%%%%%%%%
\begin{figure}[t]
\begin{center}
\vspace{0.05in} \hspace{-0.05in}{\scalebox{0.46}[0.46]{%
\includegraphics{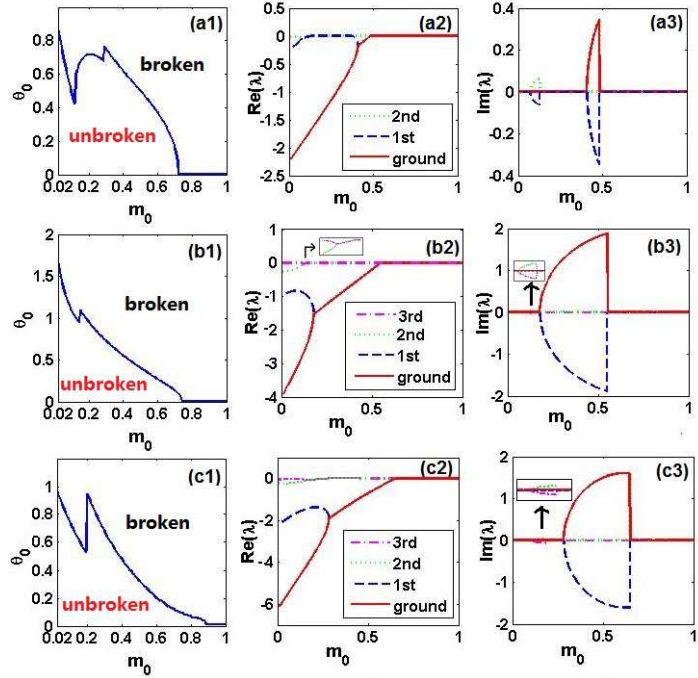}}}
\end{center}
\par
\vspace{-0.25in}
\caption{{\protect\small \textbf{Linear spectrum in $\mathcal{PT}$-symmetric
potentials.}\thinspace\ (a1, b1, c1) The $\mathcal{PT}$-symmetry of states
produced by linear eigenvalue problem (\protect\ref{ls}), with the
position-dependent diffraction coefficient (\protect\ref{sdm}) and
generalized $\mathcal{PT}$-symmetric Scarf-II potentials (\protect\ref{gpsv}%
)- (\protect\ref{gpsw}), is unbroken/broken in the domains below/above the
solid blue curves. The results are displayed for $\protect\alpha =1,2,3$,
respectively. (a2, b2, c2)  Real and (a3, b3, c3) imaginary parts of first
several eigenvalues $\protect\lambda $ as a function of $m_{0}$ for $\protect%
\theta _{0}=0.6,1,0.7$, respectively. The insets in (b2) and (b3, c3)  show,
severally, the separation and coalescence of the real and imaginary parts of
the second and third energy. }}
\label{spe1}
\end{figure}
%%%%%%%%%%%%%%%%%%%%%%%%%%%%%%%%%%%%%%%%%%%%%%%%%%%%

\vspace{0.1in} \noindent \textbf{Linear spectra of unbroken and broken $%
\mathcal{PT}$-symmetric phases.} We now introduce the diffraction
coefficient with a localized spatial modulation,
\begin{equation}
m(x)=m_{\alpha }(x)=m_{0}\mathrm{sech}^{\alpha }x+1,  \label{sdm}
\end{equation}%
and a physically relevant $\mathcal{PT}$-symmetric ingredient of the model
in the form of a generalized Scarf-II potential
\begin{eqnarray}
V_{\alpha }(x) &=&{\mathcal{V}_{1}}\,\mathrm{sech}^{2}x+{\mathcal{V}_{2}}\,%
\mathrm{sech}^{\alpha }x,\qquad \quad \,\,  \label{gpsv} \\
W_{\alpha }(x) &=&\left( \mathrm{tanh}\,x\right) \left( \mathrm{sech}%
\,x\right) \,[{\mathcal{W}_{1}}+{\mathcal{W}_{2}}\,\mathrm{sech}^{\alpha }x],
\label{gpsw}
\end{eqnarray}%
where $m_{0}>-1$ and $\alpha >0$ are both real constants, \thinspace\ ${%
\mathcal{V}_{1}}=-\frac{1}{4}(4\theta _{0}^{2}+\alpha ^{2}+6\alpha +8)$%
,\thinspace\ ${\mathcal{V}_{2}}=\frac{m_{0}}{4}(3\alpha ^{2}+8\alpha +4)$%
,\thinspace\ ${\mathcal{W}_{1}}=-\theta _{0}(\alpha +3)$, and ${\mathcal{W}%
_{2}}=-m_{0}\theta _{0}(2\alpha +3)$. Here $m_{0}$ and $\theta _{0}$ account
for the modulation of the local diffraction coefficient $m_{\alpha }$, even
real potential $V_{\alpha }$, and odd gain-loss distribution $W_{\alpha }$,
respectively. For $m_{0}=0$, Eq. (\ref{sdm}) yields a constant diffraction
coefficient, $m=1$, hence Eq.~(\ref{nls}) reduces to the usual NLS equation.
The corresponding complex-valued potential (\ref{gpsv}) and (\ref{gpsw})
becomes the usual Scarf-II potential: $U(x)={\mathcal{V}_{1}}\mathrm{sech}%
^{2}x+i{\mathcal{W}_{1}}\mathrm{sech}x\,\mathrm{tanh}x$ ~\cite{real41},
where the corresponding Hamiltonian $H(x)=-\partial _{x}^{2}+U(x)$ may have
two branches of energy eigenvalues and interpreted as the so-called
quasi-parity~\cite{real42}. Moreover, the Hamiltonian $H(x)$ can be shown to
exhibit the spontaneous $\mathcal{PT}$-symmetry breaking when the strength
of the imaginary part, $|{\mathcal{W}_{1}}|$, exceeds a threshold $-{%
\mathcal{V}_{1}}+1/4$~\cite{real4}. Below, we use the complex potential with
$m_{0}>0$ and $\alpha >0$, which may be viewed as a generalization of the
basic Scarf-II potential.

First, we consider linear spectra of phases with unbroken and broken $%
\mathcal{PT}$ symmetries in the framework of the linear spectral problem
with $\mathcal{PT}$-symmetric operator $\hat{L}$ that includes variable
diffraction coefficient (\ref{sdm}) and $\mathcal{PT}$-symmetric potential (%
\ref{gpsv})-(\ref{gpsw}). The problem is based on the equation for localized
eigenfunctions $\Phi (x)$ and respective eigenvalues $\lambda $,
\begin{equation}
\hat{L}\Phi (x)=\lambda \Phi (x),\quad \hat{L}\equiv -\frac{d}{dx}\left( m(x)%
\frac{d}{dx}\right) +V(x)+iW(x).  \label{ls}
\end{equation}%
In the limit of $m_{0}=0$ in Eq. (\ref{sdm}), $\hat{L}$ amounts to the
standard Hamiltonian operator with the usual $\mathcal{PT}$-symmetric
Scarf-II potential, which has been studied in detail by means of analytical
and numerical methods \cite{real4, yanchaos16}. For $m_{0}>0$, we focus on
natural values of powers in Eq. (\ref{sdm}), $\alpha =1,2,$ and $3$. By
means of the available numerical Fourier spectral algorithm \cite{spm, sphm}%
, we find symmetry-breaking boundaries in the $(m_{0},\theta _{0})$
parameter plane, below and above which the $\mathcal{PT}$ symmetry is
unbroken and broken, respectively, as shown in Figs.~\ref{spe1}(a1, b1,c1)).
It is seen that, for fixed $m_{0}$, there always exists a critical values of
$\theta _{0}$, beyond which the symmetry-breaking phase transition makes the
spectra complex-valued. On the other hand, for given $\theta _{0}$, the
phase transition may occur more than once with the increase of $m_{0}$,
featuring transitions of the energy spectra first from real to complex, then
back to real (\emph{restoration} of the $\mathcal{PT}$ symmetry), and
finally again from real to complex values. This scenario is drastically
different from what is known in the case of the usual Hamiltonian with the $%
\mathcal{PT}$-symmetric Scarf-II potential~\cite{real4}. To illustrate these
findings, a few lowest energy levels are displayed in Figs.~\ref{spe1}(a2,
b2, c2) (real parts) and Figs.~\ref{spe1}(a3, b3, c3) (imaginary parts).

%%%%%%%%%%%%%%%%%%%%%%%%%%%%%%%%%%%%%%%%%%%%%%%%%%%%%
\begin{figure}[!t]
\begin{center}
{\scalebox{0.75}[0.7]{\includegraphics{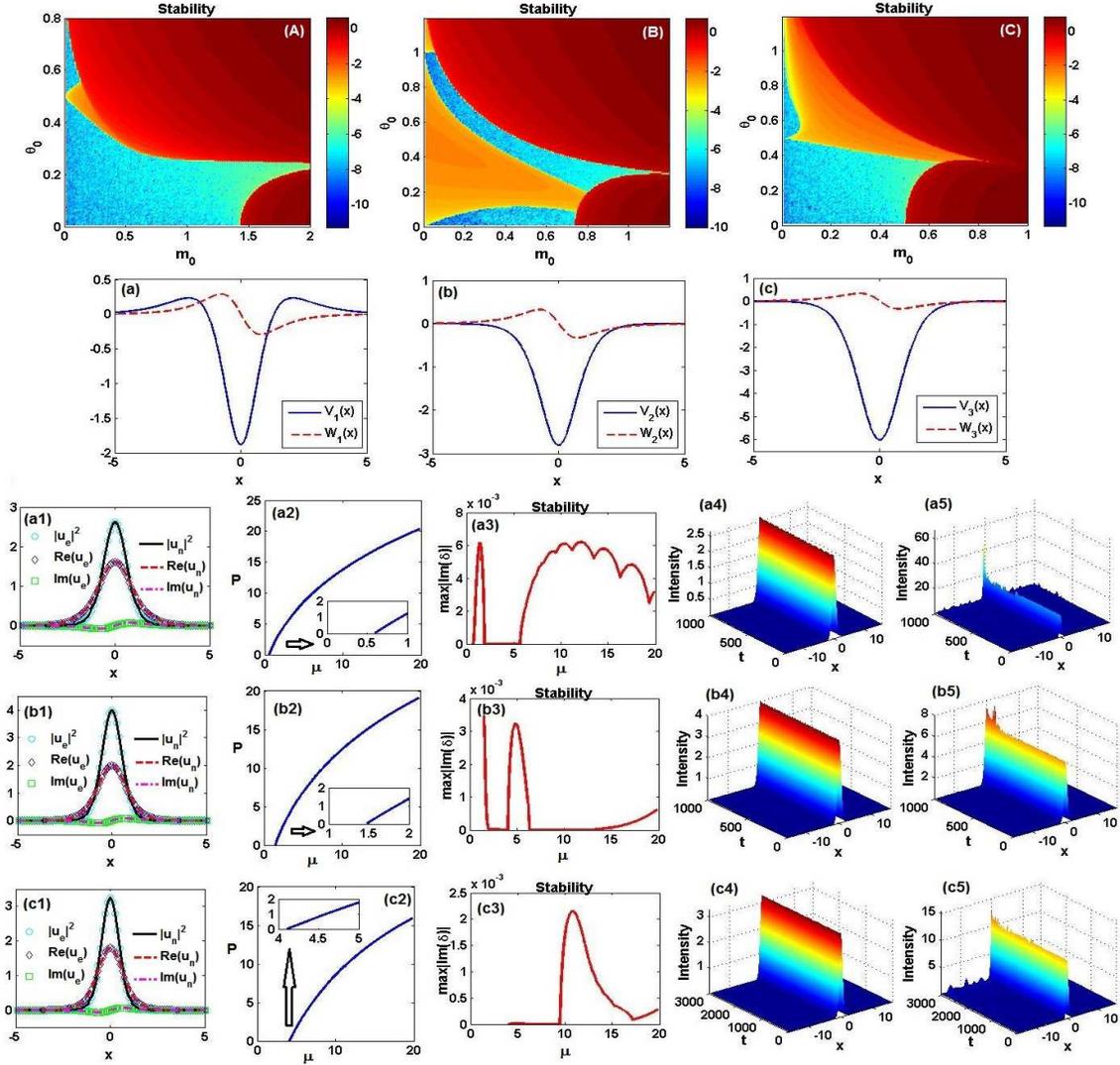}}}
\end{center}
\par
\vspace{-0.25in}
\caption{{\protect\small \textbf{Stability of exact nonlinear modes (\protect
\ref{nlm}).} (A, B, C): The linear-stability map, as obtained from the
numerical solution of Eq.~(\protect\ref{stable}), in the $(m_{0},\protect%
\theta _{0})$ plane for $\protect\alpha =1,2,3$, respectively. The stability
is determined by the largest absolute value of imaginary parts of the
linearization eigenvalue $\protect\delta $, which are displayed by means of
the color bar on a common logarithmic scale, $\mathrm{log}[\mathrm{max}(|%
\mathrm{Im}(\protect\delta )|)]$. (a, b, c) The real and imaginary parts of
the $\mathcal{PT}$-symmetric complex potentials determined by Eqs.~(\protect
\ref{gpsv}) and (\protect\ref{gpsw}) for $(\protect\alpha ,m_{0},\protect%
\theta _{0})=(1,0.5,0.1),(2,0.4,0.1),(3,0.2,0.1)$, respectively. For $%
\protect\alpha =1,2,3$, which correspond to (a1, b1, c1), respectively, the
real parts, imaginary parts, and intensity profiles are displayed, as
obtained form the exact solutions, $\protect\phi _{e}$, and their
numerically found counterparts (fundamental solitons), $\protect\phi _{n}$,
for $\protect\mu =9/4,4,25/4$, respectively. (a2, b2, c2): Power $P$ versus
the frequency, $-\protect\mu $. (a3, b3, c3): Linear-stability spectra of
numerically found fundamental solitons versus $\protect\mu $; (a4, b4, c4):
The stable evolution of the solitons (no matter exact or numerical ones, the
evolution being identical) from (a1, b1, c1). (a5, b5, c5): Unstable
evolution of the numerically found solitons at $\protect\mu =10,\protect\mu %
=5,\protect\mu =10.5$, respectively. The insets in (a2, b2, c2) indicate power
 curves near the lower cutoff points of $\mu$. }}
\label{num1}
\end{figure}

%%%%%%%%%%%%%%%%%%%%%%%%%%%%%%%%%%%%%%%%%%%%%%%%%%%%

\vspace{0.1in} \noindent \textbf{Nonlinear localized modes and their
instability.} For the given spatial profile of the diffraction coefficient (%
\ref{sdm}) and the generalized $\mathcal{PT}$-symmetric Scarf-II potential (%
\ref{gpsv})-(\ref{gpsw}), it is possible to find analytically particular
\emph{exact solutions} for stationary nonlinear localized modes (bright
solitons) of Eq.~(\ref{nls}) for $g>0$ in Eq. (\ref{nls}), i.e., the
self-focusing sign of the cubic nonlinearity (see Methods):
\begin{equation}
\psi (x,z)=\rho \,\mathrm{sech}^{\frac{\alpha +2}{2}}x\,\exp \left[ i\theta
_{0}\mathrm{tan}^{-1}(\sinh x)+i\mu z\right] ,  \label{nlm}
\end{equation}%
where $\rho =\sqrt{m_{0}(4\theta _{0}^{2}+3\alpha ^{2}+10\alpha +8)/(4g)}$,
and the propagation constant is $-\mu =-\left( \frac{\alpha }{2}+1\right)
^{2}$. Without loss of generality, we display subsequent results for the
normalization defined by $g=1$.

The integral power of the nonlinear localized modes (\ref{nlm}) is $%
P=\int_{-\infty }^{+\infty }|\psi (x,z)|^{2}dx=\rho ^{2}\int_{-\infty
}^{+\infty }\mathrm{sech}^{\alpha +2}xdx$, which is $\frac{1}{2}\pi \rho ^{2}
$, $\frac{4}{3}\rho ^{2}$, $\frac{3}{8}\pi \rho ^{2}$ for $\alpha =1,2,3$,
respectively. It is also relevant to examine the transverse power flow of
these modes (alias the Poynting vector), $S(x)=\left( i/2\right) (\phi \phi
_{x}^{\ast }-\phi ^{\ast }\phi _{x})=\rho ^{2}\theta _{0}\mathrm{sech}%
^{\alpha +3}(x)$, whose sign is solely determined by $\theta _{0}$ for any $%
\alpha >0$. It is clearly seen from Eq.~(\ref{gpsw}) that the signs of
gain-loss distribution $W_{\alpha }$ are also determined by the single
parameter $\theta _{0}$, for $m_{0}\geq 0$ and $\alpha >0$. Thus we conclude
that the power always flows from the gain region to one carrying the loss,
regardless of the sign of $\theta _{0}$.

For different powers $\alpha =1,2,3$, we aim to study the linear stability
of exact nonlinear modes (\ref{nlm}) by numerically calculating the largest
absolute value of imaginary parts of the linearization eigenvalue $\delta $
from Eq.~(\ref{stable}), see below, in the modulation-parameter plane $%
(m_{0},\theta _{0})$. Figures~\ref{num1}(A,B,C) exhibit the so found
stability (blue) and instability (other colors) regions of the localized
modes (\ref{nlm}) for $\alpha =1,2,3$, respectively. For these three cases,
we, respectively, choose three stable points (i.e., ones with the unbroken $%
\mathcal{PT}$ symmetry): $(m_{0},\theta _{0})=(0.5,0.1),(0.4,0.1),(0.2,0.1)$%
, to display the corresponding $\mathcal{PT}$-symmetric potentials in Figs.~%
\ref{num1}(a,b,c). The single-well potential becomes deeper, whereas the
amplitude of the gain-and-loss distribution slowly decreases, as $\alpha $
increases and $m_{0}$ decreases. For these three cases, we show exact
solitonic solutions (\ref{nlm}) and their numerically found counterparts in
Figs.~\ref{num1}(a1,b1,c1), to corroborate that they are mutually identical.
For other values of propagation constant $-\mu $, exact analytical solutions
are not available, but we can use the numerical method (validated by the
comparison with the exact solutions) to produce fundamental solitons. It
follows from Figs.~\ref{num1}(a2,b2,c2) that the solitons' powers are
monotonously increasing functions of $\mu $. While existence ranges of the
numerically found solitons have nearly the same upper cutoff $\mu _{\mathrm{%
up}}\approx 19.8$, the lower cutoffs are different: $\mu _{1\mathrm{low}%
}=0.6,~\mu _{2\mathrm{low}}=1.48,~\mu _{3\mathrm{low}}=4.11$, respectively,
with a trend to growth. In the existence range of the numerically found
solitons, the dependence of the corresponding linear-stability eigenvalues
(the largest absolute value of the imaginary part of the perturbation growth
rate $\delta $) on the propagation constant $-\mu $ is shown in Figs.~\ref%
{num1}(a3,b3,c3). It is seen that the numerically found soliton solutions in
Figs.~\ref{num1}(a1,b1,c1), corresponding to $\mu =2.25,4, 6.25$,
respectively, are completely stable, in accordance with the stability of the
corresponding exact nonlinear modes (see Figs.~\ref{num1}(A,B,C)).

To validate the linear-stability results, we have simulated the propagation,
by taking the input provided by the stationary modes in Figs.~\ref{num1}%
(a1,b1,c1) with the addition of $2\%$ random perturbations, as is shown in
Figs.~\ref{num1}(a4,b4,c4), respectively. In practice, we simulate the beam
propagation with the initial input $\psi (x,z=0)=\phi (x)(1+\epsilon )$,
where $\phi (x)$ is a nonlinear mode, and $\epsilon $ is a complex broadband
random perturbation. In MATLAB, the $2\%$ white noise $\epsilon $ can be
realized by utilizing a random matrix such as $\epsilon =0.04(\mathrm{rand}%
(N,1)-0.5)(1+i)/\sqrt{2}$, where $\mathrm{rand}(N,1)$ is an $N-by-1$ array
of pseudorandom uniform values on the open interval $(0,1)$ (similarly in
other cases, even for the 2D situation). Finally, based on the linear
stability results presented in Figs.~\ref{num1}(a3,b3,c3), we choose
solutions that are predicted to be unstable (with $\mu =10, 5, 10.5$), to
test the dynamical behavior of the corresponding numerically found soliton
solutions. It is found that they are indeed (weakly) unstable, see Figs.~\ref%
{num1}(a5,b5,c5)).

%%%%%%%%%%%%%%%%%%%%%%%%%%%%%%%%%%%%%%%%%%%%%%%%%%%%%
\begin{figure}[!t]
\begin{center}
\vspace{0.05in} {\scalebox{0.55}[0.5]{\includegraphics{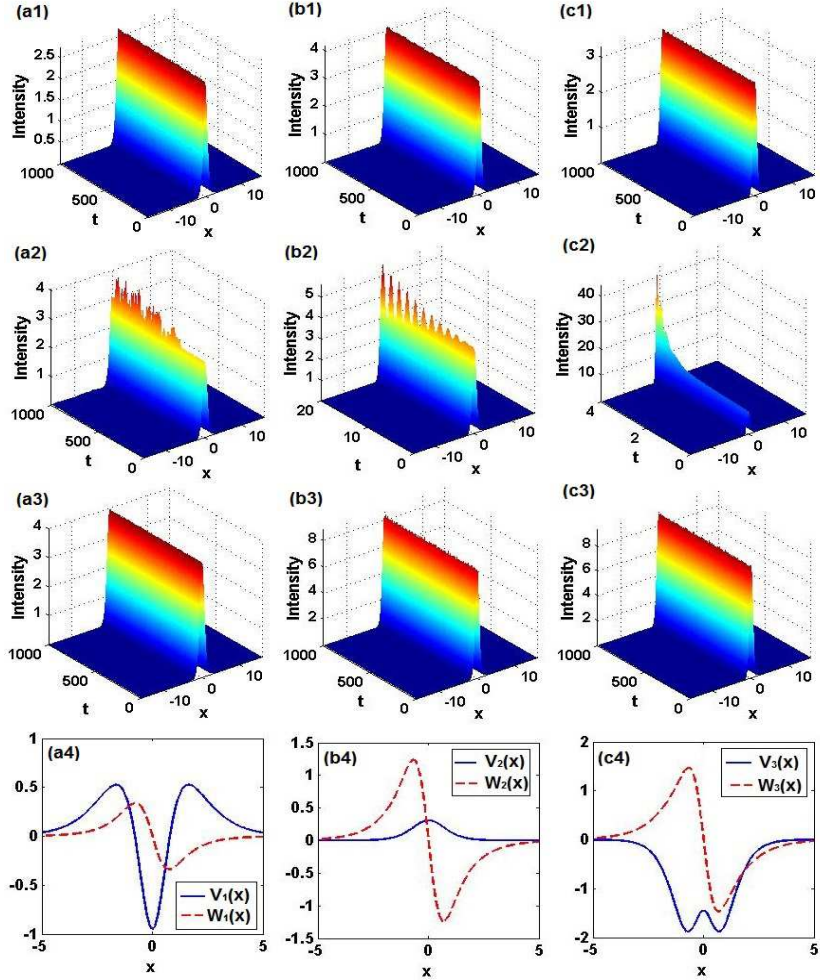}}}
\end{center}
\par
\vspace{-0.25in}
\caption{{\protect\small \textbf{Propagation dynamics of exact nonlinear
localized modes (\protect\ref{nlm})}. (a1): $m_{0}=0.5,\protect\theta %
_{0}=0.3$, unbroken; (a2): $m_{0}=0.5,\protect\theta _{0}=0.32$, unbroken;
(a3): $m_{0}=0.75,\protect\theta _{0}=0.1$, broken; (b1): $m_{0}=0.4,\protect%
\theta _{0}=0.5$, unbroken; (b2): $m_{0}=0.4,\protect\theta _{0}=0.55$,
broken; (b3): $m_{0}=0.8,\protect\theta _{0}=0.3$, broken; (c1) $m_{0}=0.2,%
\protect\theta _{0}=0.36$, unbroken; (c2): $m_{0}=0.54,\protect\theta %
_{0}=0.1$, unbroken; (c3): $m_{0}=0.54,\protect\theta _{0}=0.36$, broken.
(a4,b4,c4): Real and imaginary parts of the $\mathcal{PT}$-symmetric
potential corresponding to cases (a3,b3,c3), respectively. Other parameters
are, severally, $\protect\alpha =1,2,$ and $3$, for the first, second, and
third columns. }}
\label{evo1}
\end{figure}
%%%%%%%%%%%%%%%%%%%%%%%%%%%%%%%%%%%%%%%%%%%%%%%%%%%%

%%%%%%%%%%%%%%%%%%%%%%%%%%%%%%%%%%%%%%%%%%%%%%%%%%%%%
\begin{figure}[!t]
\begin{center}
\vspace{0.05in} {\scalebox{0.56}[0.5]{\includegraphics{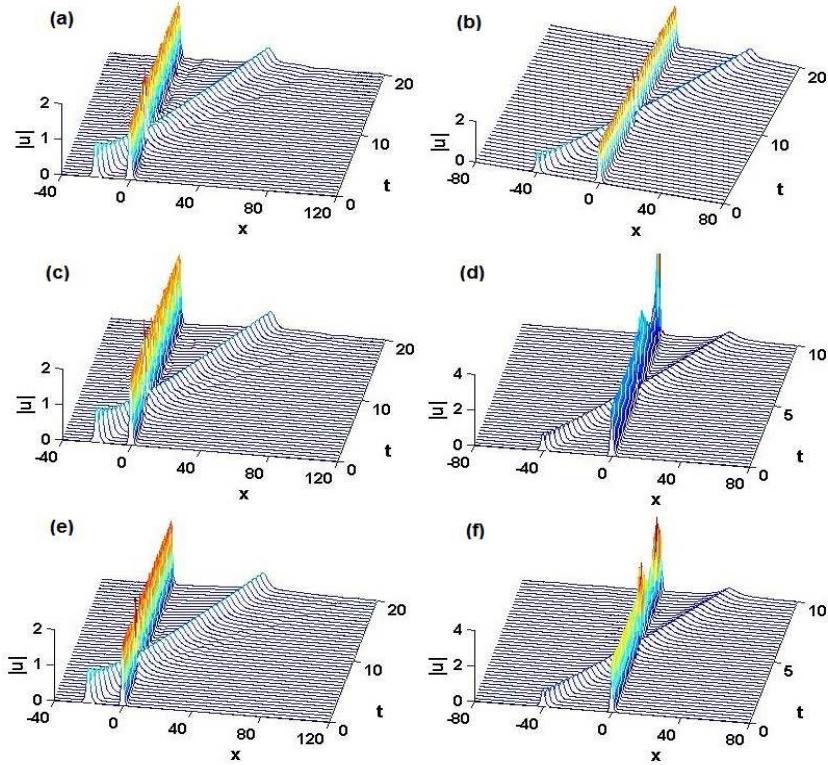}}}
\end{center}
\par
\vspace{-0.25in}
\caption{{\protect\small \textbf{Collisions between exact bright solitons
and boosted sech-shaped solitary pulses, produced by simulations of Eq.~(%
\protect\ref{nls})}. (a) $\protect\alpha =1,m_{0}=0.5,\protect\theta %
_{0}=0.1 $ (unbroken), with the input composed of the exact soliton (\protect
\ref{nlm}) and the solitary pulse $\mathrm{sech}(x+20)e^{1.8ix}$, (b) $%
\protect\alpha =1,m_{0}=0.75,\protect\theta _{0}=0.1$ (broken), with the
initial solitary pulse $\mathrm{sech}(x+40)e^{2.2ix}$; (c) $\protect\alpha %
=2,m_{0}=0.4,\protect\theta _{0}=0.1$ (unbroken), with the initial solitary
pulse $\mathrm{sech}(x+20)e^{1.8ix}$, (d) $\protect\alpha =2,m_{0}=0.8,%
\protect\theta _{0}=0.3$ (broken), with the initial solitary pulse $\mathrm{%
sech}(x+40)e^{4ix}$; (e) $\protect\alpha =3,m_{0}=0.2,\protect\theta %
_{0}=0.1 $ (unbroken), with the initial solitary pulse $\mathrm{sech}%
(x+20)e^{1.8ix}$, (f) $\protect\alpha =3,m_{0}=0.54,\protect\theta _{0}=0.36$
(broken), with the initial solitary pulse $\mathrm{sech}(x+40)e^{4ix}$.
Other parameters are $\protect\alpha =1,2,3$ for the first, second, and
third row, respectively.}}
\label{col1}
\end{figure}
%%%%%%%%%%%%%%%%%%%%%%%%%%%%%%%%%%%%%%%%%%%%%%%%%%%%

%%%%%%%%%%%%%%%%%%%%%%%%%%%%%%%%%%%%%%%%%%%%%%%%%%%%%
\begin{figure}[!t]
\begin{center}
\vspace{0.05in} {\scalebox{0.6}[0.6]{\includegraphics{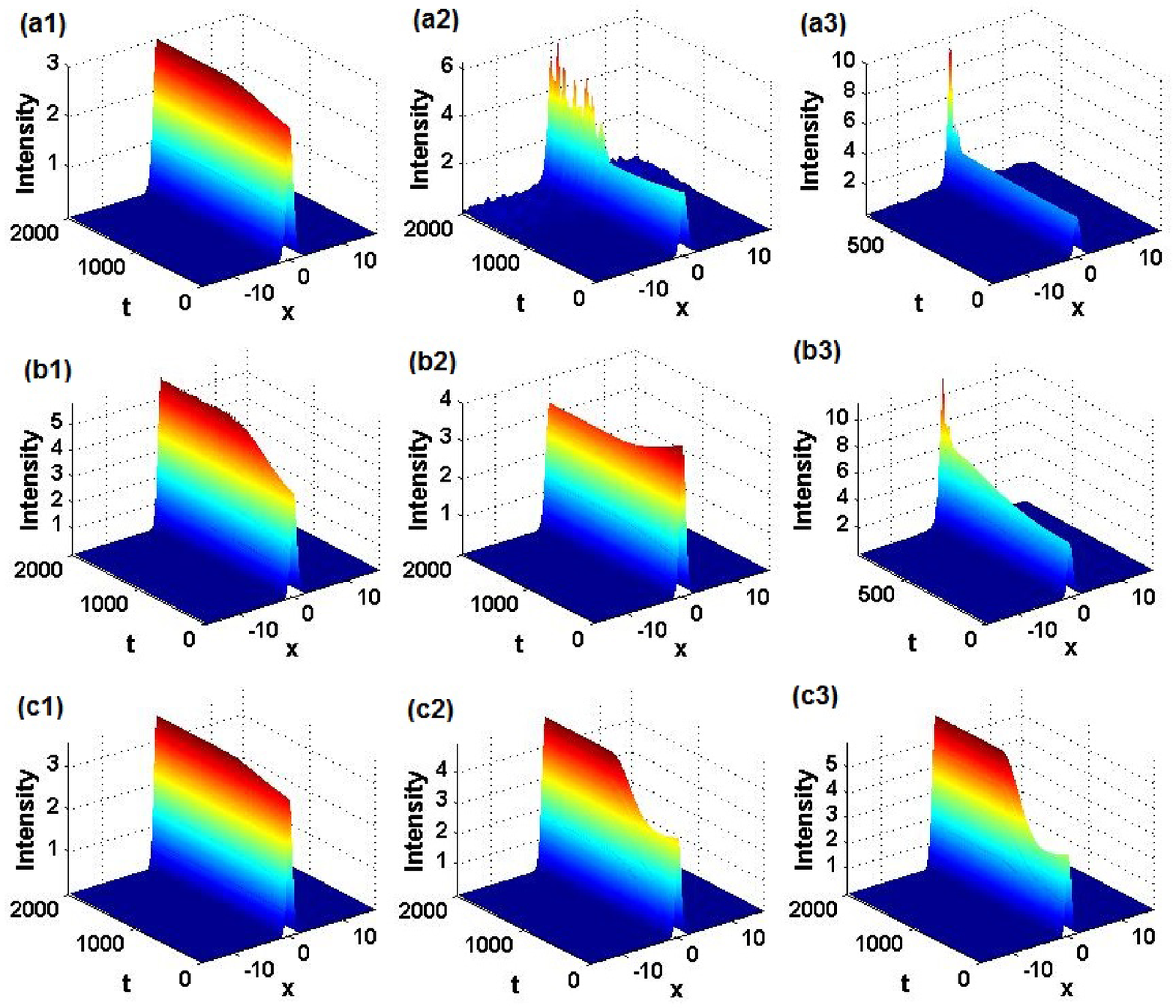}}}
\end{center}
\par
\vspace{-0.25in}
\caption{{\protect\small \textbf{The transformation of initially stable
nonlinear modes, produced by simulations of Eq.~(\protect\ref{tnls})}. (a1) $%
m_{01}=0.5,\protect\theta _{01}=0.1,\protect\theta _{02}=0.3$, (a2) $%
m_{01}=0.5,m_{02}=0.75,\protect\theta _{01}=0.1$, (a3) $%
m_{01}=0.5,m_{02}=0.75,\protect\theta _{01}=0.1,\protect\theta _{02}=0.3$;
(b1) $m_{01}=0.4,\protect\theta _{01}=0.1,\protect\theta _{02}=0.5$, (b2) $%
m_{01}=0.4,m_{02}=0.6,\protect\theta _{01}=0.1$, (b3) $m_{01}=0.4,m_{02}=0.6,%
\protect\theta _{01}=0.1,\protect\theta _{02}=0.5$; (c1) $m_{01}=0.2,\protect%
\theta _{01}=0.1,\protect\theta _{02}=0.36$, (c2) $m_{01}=0.2,m_{02}=0.54,%
\protect\theta _{02}=0.36$, (c3) $m_{01}=0.2,m_{02}=0.54,\protect\theta %
_{01}=0.1,\protect\theta _{02}=0.36$. Cases (a1,b1,b2,c1) exhibit the stable
transformation between parameter sets correspond to stable linear states
with unbroken $\mathcal{PT}$ symmetry. (c2,c3): Stable transformation
between parameter sets corresponding to unbroken and broken $\mathcal{PT}$%
-symmetry of the linear states. (a2,a3,b3): Unstable transformation of the
initially stable nonlinear mode between parameter sets corresponding to
unbroken and broken $\mathcal{PT}$-symmetry of the linear states. Other
parameters are $\protect\alpha =1,2,3$ for the first, second, and third row,
respectively. }}
\label{exc1}
\end{figure}
%%%%%%%%%%%%%%%%%%%%%%%%%%%%%%%%%%%%%%%%%%%%%%%%%%%%

Further, following the results of the linear-stability analysis shown in
Figs.~\ref{num1}(A,B,C), we have tested the propagation dynamics of the
exact nonlinear modes (\ref{nlm}), varying the parameter $m_{0}$ or $\theta
_{0}$ in Eqs. (\ref{sdm}) - (\ref{gpsw}). For the brevity's sake, hereafter
we use the words ``unbroken" and ``broken" to mention if the corresponding
linear modes, considered above, do or do not keep their $\mathcal{PT}$
symmetry. For fixed $\alpha =1$ and $m_{0}=0.5$, as $\theta _{0}$ increases
continuously from $0.1$ to $0.3$ (unbroken), nonlinear modes (\ref{nlm})
always feature stable propagation dynamics, as shown in Fig.~\ref{evo1}(a1).
However, as $\theta _{0}$ increases to slightly larger values, e.g., $0.32$
(unbroken), instability sets in, see Fig.~\ref{evo1}(a2). Moreover, we have
found that a \emph{stable} nonlinear localized mode (\ref{nlm}) with $%
(m_{0},\theta _{0})=(0.75,0.1)$ belongs to the region of \emph{broken}
linear $\mathcal{PT}$-symmetry (see Fig.~\ref{evo1}(a3)), with the real part
of the corresponding complex potential having a single-well shape, see Fig.~%
\ref{evo1}(a4). The latter result implies that the exact $\mathcal{PT}$%
-symmetric nonlinear modes may be stable while their linear counterparts are
not.

For fixed $\alpha =2$ and $m_{0}=0.4$, as $\theta _{0}$ grows continuously
from $0.1$ to $0.5$ (unbroken) and further to $0.55$ (broken), similar
results occur, see Figs.~\ref{evo1}(b1,b2). Here too, we find a stable
nonlinear localized mode, while the $\mathcal{PT}$ symmetry of its linear
counterpart is broken, for parameters $(m_{0},\theta _{0})=(0.8,0.3)$, see
Fig.~\ref{evo1}(b3), even if the real part of the corresponding complex
potential is a weak barrier (rather than a well), whose amplitude is smaller
than that of the corresponding gain-loss distribution, see Fig.~\ref{evo1}%
(b4). For $\theta _{0}=0.1$, we increased $m_{0}$ continuously from $0.4$ to
$0.6$ (unbroken), so that a continuous family of stable solitons could also
be readily found.

For $\alpha =3$, when we fix $m_{0}=0.2$ and increase $\theta _{0}$
continuously from $0.1$ to $0.36$ (unbroken), similar results still hold,
see Fig.~\ref{evo1}(c1). On the other hand, when, for $\theta _{0}=0.1$, $%
m_{0}$ increases to $0.54$, the nonlinear localized mode may be unstable, as
in Fig.~\ref{evo1}(c2), even if the $\mathcal{PT}$ symmetry of the linear
state remains unbroken. Most interesting, if we increase $\theta _{0}$ to $%
0.36$ in the same case, the nonlinear localized mode \emph{restabilizes
itself}, while the $\mathcal{PT}$ symmetry of the linear state gets broken,
see Fig.~\ref{evo1}(c3). Another interesting feature is that the real part
of the corresponding complex potential may now exhibit a double-well shape,
as shown in Fig.~\ref{evo1}(c4).

\vspace{0.1in} \noindent \textbf{Interactions between exact bright solitons
and sech pulses.} To additionally test the robustness of the exact bright
solitons (\ref{nlm}), we simulated their collisions with boosted (moving)
sech-shaped solitary pulses. For $\alpha =1,2,3$, we respectively choose $%
(m_{0},\theta _{0})=(0.5,0.1),(0.4,0.1),(0.2,0.1)$ with the unbroken $%
\mathcal{PT}$ symmetry of the linear state and consider initial conditions $%
\psi _{\alpha }(x,0)=\phi _{\alpha }(x)+\mathrm{sech}(x+20)e^{1.8ix}$ with $%
\phi _{\alpha }(x)$ given by Eq.~(\ref{nlm}), and $e^{1.8ix}$ imposing the
boost onto the sech pulse with the unitary amplitude and initial position at
$x=-20$. Direct simulations demonstrate that the exact nonlinear modes $\phi
_{\alpha }(x)$ propagate steadily without any change of shape and velocity
after the collision, see Figs.~\ref{col1}(a,c,e). In other simulations, we
chose values $(m_{0},\theta _{0})=(0.75,0.1),(0.8,0.3),(0.54,0.36)$, which
correspond to the linear modes with broken $\mathcal{PT}$ symmetry, and took
the corresponding initial conditions as $\psi _{1}(x,0)=\phi _{1}(x)+\mathrm{%
sech}(x+40)e^{2.2ix}$ with $\phi _{1}(x)$ given by Eq.~(\ref{nlm}) for $%
\alpha =1$, as well as $\psi _{2,3}(x,0)=\phi _{2,3}(x)+\mathrm{sech}%
(x+40)e^{4ix}$, with $\phi _{2,3}(x)$ given by Eq.~(\ref{nlm}) for $\alpha
=2,3$. In the case of $\alpha =1$, the exact soliton still demonstrates the
steady propagation after the collision, while in the cases of $\alpha =2,3$
the amplitude of the initial exact soliton rapidly grows after the
collision, thus manifesting interaction-induced instability.

\vspace{0.1in} \noindent\textbf{Adiabatic transformation of stable nonlinear
modes.} Here, we elaborate three different scenario of dynamical control of
nonlinear localized modes, making use of adiabatically varying parameters of
the potential, $m_{0}\rightarrow m_{0}(z)$ or $\theta _{0}\rightarrow \theta
_{0}(z)$, cf. Refs.~\cite{t10, yansr16,yanchaos16,yanstable2017}. We choose the following
temporal-modulation pattern:%
\begin{equation}
\left\{ m_{0},\theta _{0}\right\} (z)=%
\begin{cases}
\left( \left\{ m_{02},\theta _{02}\right\} -\left\{ m_{01},\theta
_{01}\right\} \right) \sin (\pi z/2000)+\left\{ m_{01},\theta _{01}\right\} ,
& \text{$0\leq z<1000$}, \\
\left\{ m_{02},\theta _{02}\right\} , & \text{$z\geq 1000.$}%
\end{cases}
\label{excite-s}
\end{equation}%
This implies replacing Eq. (\ref{nls}) by
\begin{equation}
i\frac{\partial \psi }{\partial z}=\left[ -\frac{\partial }{\partial x}m(x,z)%
\frac{\partial }{\partial x}+V(x,z)+iW(x,z)-g|\psi |^{2}\right] \psi ,
\label{tnls}
\end{equation}%
where $m(x,z),V(x,z),W(x,z)$ are respectively given by Eqs.~[(\ref{sdm}), (%
\ref{gpsv}), (\ref{gpsw})] with $m_{0}\rightarrow m_{0}(z)$ and $\theta
_{0}\rightarrow \theta _{0}(z)$.

For the cases of $\alpha =1,2,3$, Figs.~\ref{exc1}(a1,b1,c1), respectively,
exhibit stable switch of the nonlinear localized modes $\psi (x,z)$ governed
by Eq.~(\ref{tnls}), starting from the initial condition given by Eq.~(\ref%
{nlm}). In these figures, we demonstrate the transformation of initially
stable nonlinear localized modes for $(m_{01},\theta
_{01})=(0.5,0.1),(m_{01},\theta _{01})=(0.4,0.1),(m_{01},\theta
_{01})=(0.2,0.1)$, with $\alpha =1,2,3$ to ones corresponding, respectively,
to $(m_{02},\theta _{02})=(0.5,0.3),(m_{02},\theta
_{02})=(0.4,0.5),(m_{02},\theta _{02})=(0.2,0.36)$, i.e., $m_{0}$ is fixed,
while $\theta _{0}$ varies. In these cases, both the initial and final
parameter values correspond to linear modes with unbroken $\mathcal{PT}$
symmetry. Likewise, fixing $\theta _{0}$ and setting $m_{0}\rightarrow
m_{0}(z)$ as per by Eq.~(\ref{excite-s}), we can perform similar
transformations of stable exact nonlinear localized modes, as is shown in
Figs.~\ref{exc1}(a2,b2,c2).  Figure~\ref{exc1}(b2) displays a typical stable
transformation of the nonlinear modes (\ref{nlm}) with $\alpha =2$, also
ending with parameters corresponding to the linear state with unbroken $%
\mathcal{PT}$ symmetry.

On the other hand, Fig.~\ref{exc1}(a2) reveals unstable transformation of
the nonlinear modes (\ref{nlm}) with $\alpha =1$, and Fig.~\ref{exc1}(c2)
exhibits stable transformation of the nonlinear modes (\ref{nlm}) with $%
\alpha =3$, both from initial parameters corresponding to the linear state
with unbroken $\mathcal{PT}$ symmetry to final ones corresponding to broken $%
\mathcal{PT}$ symmetry in the linear state. Finally, the transformation is
also implemented by varying both $m_{0}$ and $\theta _{0}$. Instability of
such simultaneous transformation, observed in Fig.~\ref{exc1}(a3) may be
attributed to the instability of the corresponding single-parameter
variation, shown above in Fig.~\ref{exc1}(a2). Further, Fig.~\ref{exc1}(b3)
reveals that the simultaneous variation of $m_{0}$ and $\theta _{0}$ may
give rise to an unstable output, while both corresponding the
single-component variations are stable, see Figs.~\ref{exc1}(b1,b2).
Nevertheless, in Fig.~\ref{exc1}(c3) we stably transform an initially exact
nonlinear mode, with unbroken $\mathcal{PT}$ symmetry of the respective
linear state, to another exact nonlinear mode, which corresponds to the
broken $\mathcal{PT}$ symmetry in the linear state, varying both parameters.
The stable outcome of the simultaneous variation of $m_{0}$ and $\theta _{0}$
may be possible if the separate variation of each parameter produced a
stable output.

Thus, we conclude that the simultaneous variation of the two parameters
leads to an unstable output if either of the two corresponding separate
variations is unstable, see Figs.~\ref{exc1}(a1, a2, a3). If both separate
excitations produce stable outputs, their simultaneous action may give
either an unstable output, see Figs.~\ref{exc1}(b1,b2,b3), or a stable one,
see Figs.~\ref{exc1}(c1,c2,c3)).

%%%%%%%%%%%%%%%%%%%%%%%%%%%%%%%%%%%%%%%%%%%%%%%%%%%%%
\begin{figure}[t]
\begin{center}
{\scalebox{0.56}[0.56]{\includegraphics{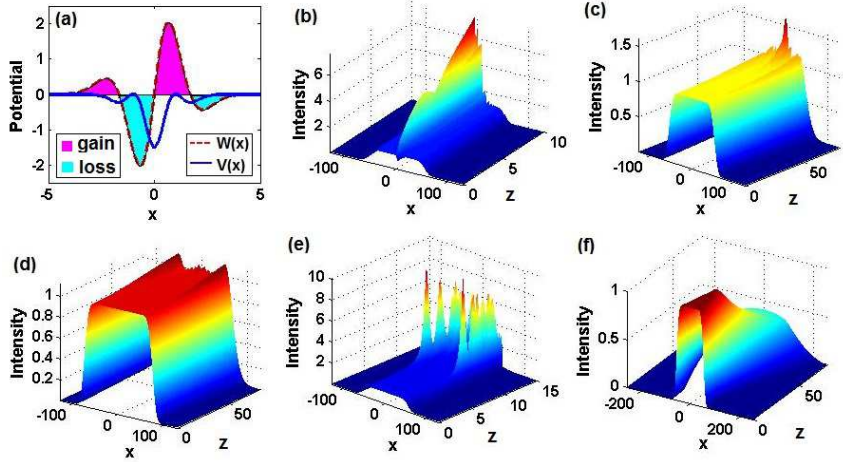}}}
\end{center}
\par
\vspace{-0.25in}
\caption{{\protect\small \textbf{Constant-intensity waves in linear and
nonlinear $\mathcal{PT}$-symmetric systems with the complex potential of
Hermite-Gauss (HG) type}. (a) Real (solid blue) and imaginary (dashed red)
parts of the complex potential $V(x)+iW(x)$ given by Eq. (\protect\ref{v})
(hereafter, magenta- and cyan-filled areas designate the presence of the
gain and loss, respectively. (b) The evolution of the constant-amplitude
wave without the correctly introduced initial phase at $z=0$ in the linear
setting (}$g=0$){\protect\small . Spatial diffraction of (c) narrow and (d)
wide truncated constant-intensity waves. The evolution of the
constant-amplitude wave with the correct phase at } ${\protect\small z=0}$
{\protect\small under (e)} {\protect\small the self-focusing ($g=1$)
nonlinearity and (f) the self-defocusing ($g=-1$) nonlinearity.}}
\label{num-ci1}
\end{figure}
%%%%%%%%%%%%%%%%%%%%%%%%%%%%%%%%%%%%%%%%%%%%%%%%%%%%

%%%%%%%%%%%%%%%%%%%%%%%%%%%%%%%%%%%%%%%%%%%%%%%%%%%%%
\begin{figure}[t]
\begin{center}
{\scalebox{0.56}[0.56]{\includegraphics{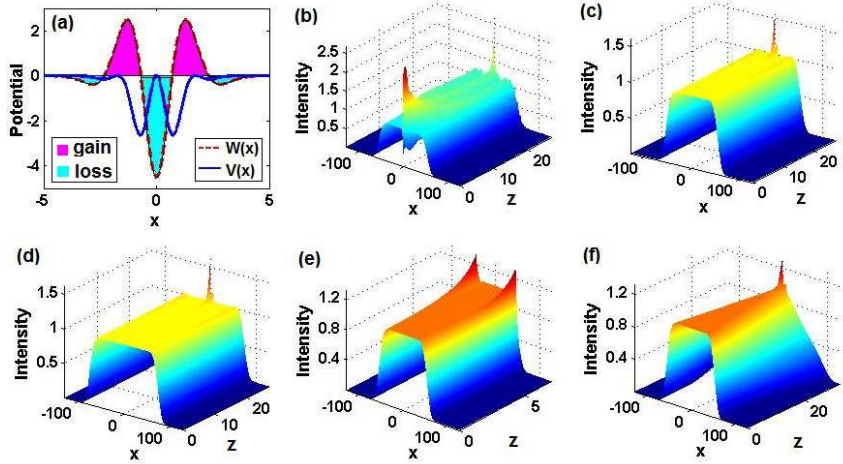}}}
\end{center}
\par
\vspace{-0.25in}
\caption{{\protect\small \textbf{Constant-intensity waves in linear and
nonlinear non-$\mathcal{PT}$-symmetric systems with the complex potential of
HG type}. The same as in Fig. \protect\ref{num-ci1}, but for
the non-}$ \mathcal{PT}${\protect\small -symmetric complex potential of the
HG type, see the text.}}
\label{num-ci2}
\end{figure}
%%%%%%%%%%%%%%%%%%%%%%%%%%%%%%%%%%%%%%%%%%%%%%%%%%%%

%%%%%%%%%%%%%%%%%%%%%%%%%%%%%%%%%%%%%%%%%%%%%%%%%%%%%
\begin{figure}[t]
\begin{center}
{\scalebox{0.56}[0.56]{\includegraphics{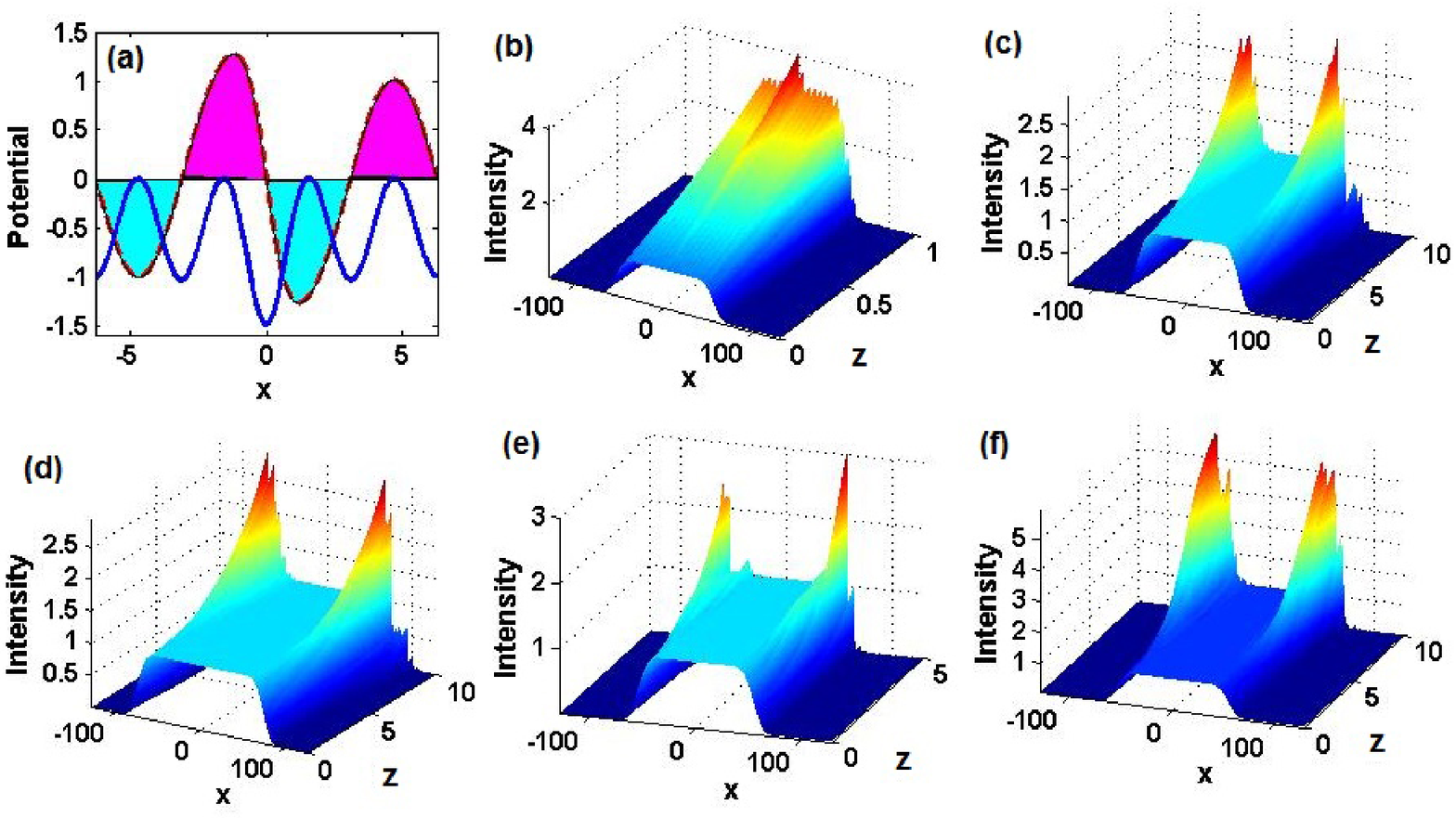}}}
\end{center}
\par
\vspace{-0.25in}
\caption{{\protect\small \textbf{Constant-intensity waves in linear and
nonlinear $\mathcal{PT}$-symmetric systems with the complex potential of $\cos$
 type}. The same as in Figs. \protect\ref{num-ci1} and
\protect\ref{num-ci2}, but for the }$\mathcal{PT}${\protect\small -symmetric
complex potential of the }$\cos ${\protect\small \ type, see the text.}}
\label{num-ci3}
\end{figure}
%%%%%%%%%%%%%%%%%%%%%%%%%%%%%%%%%%%%%%%%%%%%%%%%%%%%

\vspace{0.1in} \noindent \textbf{1D constant-intensity waves.} Recently, a
class of complex potentials that are more general than the $\mathcal{PT}$%
-symmetric complex-valued ones has been put forward~\cite%
{t11,Abdullaev,Lisbon,Vermont}. Similar to the $\mathcal{PT}$-symmetric
potentials, these more general complex potentials admit the existence of
\emph{continuous} \emph{families} of stationary states, supported by the
balance between gain and loss (unlike \emph{isolated} stationary solutions
found in generic dissipative systems), a part of which may be stable; see
Refs. \cite{t11,Abdullaev,Lisbon,Vermont}. The real and imaginary parts of
these potentials are defined as~\cite{wadati} $V(x)=-v^{2}(x)$, $%
W(x)=dv(x)/dx$, where $v(x)$ is an arbitrary real function. Here, we address
a generalization of such potentials in the model with the variable
diffraction coefficient, $m(x)$, in the form of
\begin{equation}
V(x)=-m(x)v^{2}(x),\qquad W(x)=\frac{d(m(x)v(x))}{dx},  \label{v}
\end{equation}%
where $v(x)$ is a known real function of space. If $m(x)$ (e.g., that
defined in Eq.~(\ref{sdm})) and $v(x)$ are both even functions, then the
complex potential given by Eq.~(\ref{v}) is a $\mathcal{PT}$-symmetric one.
In a more general case, when the potential is not $\mathcal{PT}$-symmetric,
it nevertheless provides for the global balance between the gain and loss,
in the case of localized or periodic functions $m(x)v(x)$, because $W(x)$ is
defined in Eq. (\ref{v}) as a full derivative.

For the general potential taken as per Eq. (\ref{v}), stationary
constant-intensity (alias CW, i.e., continuous-wave) solutions of Eq.~(\ref%
{nls}) with any $g$ are found in the form of (see Methods)
\begin{equation}
\psi (x,z)=C\exp \left[ i\int_{0}^{x}v(x)dx+igC^{2}z\right] .  \label{C}
\end{equation}%
where $C$ is a real constant amplitude. The power flow (the Poynting vector)
corresponding to the CW solution (\ref{C}) is $S(x)=C^{2}v(x)$. To examine
dynamical properties of the CW solution (\ref{C}), we consider three
different types of the complex potentials~(\ref{v}), in which the
diffraction coefficient $m(x)$ is chosen as in Eq.~(\ref{sdm}), and $v(x)$
is taken as a Hermite-Gauss (HG) function, $H_{n}(x)e^{-\omega x^{2}/2}$ ($%
H_{n}(x)$ is the Hermite polynomial, and $\omega >0$ is a frequency), or as
a simple periodic one, $v_{0}\mathrm{cos}(x)$. Generic examples of this
class of complex potentials correspond to $m(x)=0.5\mathrm{sech}(x)+1$, with
$v(x)=H_{2}(x)e^{-x^{2}/2}$ (a $\mathcal{PT}$-symmetric HG form), or $%
v(x)=H_{3}(x)e^{-x^{2}/2}$ (a non-$\mathcal{PT}$-symmetric HG form), or,
finally, $v(x)=\mathrm{cos}(x)$ (a $\mathcal{PT}$-symmetric periodic form).
The profiles and gain-loss regions of these three complex potentials~are
displayed in Figs.~\ref{num-ci1}a, \ref{num-ci2}(a), \ref{num-ci3}(a),
respectively. In the linear limit ($g=0$) in these three cases, if the CW
inputs are taken merely as $\psi (x,0)=C$, without the correct phase given
by Eq.~(\ref{C}), the beams grow fast in the center, leading to instability,
as shown in Figs.~\ref{num-ci1}(b), \ref{num-ci2}(b), and \ref{num-ci3}(b).
However, if the inputs are taken as solution~(\ref{C}) with the correct
phase, the growth of perturbations is initially suppressed, occurring later
as the modulational instability of the CW (Figs.~\ref{num-ci1}(c), \ref%
{num-ci2}c, and \ref{num-ci3}(c). It is worthy to note that the beam in the $%
\mathcal{PT}$-symmetric system with the HG complex potential steadily
propagates farther than that in the case of the non-$\mathcal{PT}$ HG
complex potential, as the $\mathcal{PT}$-symmetric potential naturally
provides for a better balance of the gain and loss. When the truncation
length of the input CW becomes larger, the stable-propagation distance
increases too under the action of the $\mathcal{PT}$-symmetric HG potential
(see Fig.~\ref{num-ci1}d), but it does not increase in the case of the non-$%
\mathcal{PT}$-symmetric HG potential (see Fig.~\ref{num-ci2}d).

We have also investigated the nonlinear evolution of the CW input in the
presence of the self-focusing and self-defocusing Kerr nonlinearity ($g=+1$
and $-1$, respectively) for the same three complex potentials. As a result,
we find that the CW state maintains stable propagation over a smaller
distance than that in the corresponding linear model, as can be seen in
Figs.~\ref{num-ci1}e, \ref{num-ci2}e, and \ref{num-ci3}e for the
self-focusing nonlinearity, and in Figs.~\ref{num-ci1}(f), \ref{num-ci2}(f),
and \ref{num-ci3}(f) for the self-defocusing nonlinearity.

\vspace{0.1in} \noindent \textbf{2D $\mathcal{PT}$-symmetric nonlinear waves.%
} Multidimensional spatial solitons are a subject of great interest to
nonlinear optics \cite{opso,op2,op2a,op2b}. We here consider the formation
of bright spatially localized solitons in the 2D $\mathcal{PT}$-symmetric
setting. In this case, the field evolution is governed by the 2D NLS
equation with a $\mathcal{PT}$-symmetric potential:
\begin{equation}
i\frac{\partial \psi }{\partial z}=\left\{ -\nabla \left[ \mathbf{m}%
(x,y)\nabla \right] +V(x,y)+iW(x,y)-g|\psi |^{2}\right\} \psi ,
\label{2D-NLS}
\end{equation}%
where $\nabla $ is the 2D gradient $(\partial _{x},\partial _{y})$, and $%
\mathbf{m}(x,y)$ is a $2\times 2$ matrix function, which we take in the
diagonal form, $m_{\alpha }(x,y)=\mathrm{diag}[m_{\alpha }(x),m_{\alpha
}(y)] $ with $m_{\alpha }(x)$ given by Eq.~(\ref{sdm}). We look for
stationary 2D modes in the form of $\psi =\phi (x,y)e^{i\mu z}$ with $\phi
(x,y)=u(x,y)e^{i\theta (x,y)}$, where $u(x,y)$ and $\theta (x,y)$ are real
amplitude and phase, respectively, which satisfy stationary equations:
\begin{gather}
(m_{\alpha }(x)u_{x})_{x}+(m_{\alpha }(y)u_{y})_{y}-[\theta
_{x}^{2}m_{\alpha }(x)+\theta _{y}^{2}m_{\alpha }(y)]u-V(x,y)u-\mu
u+gu^{3}=0,  \label{2dsys1} \\
(m_{\alpha }(x)\theta _{x}u^{2})_{x}+(m_{\alpha }(y)\theta
_{y}u^{2})_{y}-W(x,y)u^{2}=0.  \label{2dsys2}
\end{gather}%
2D $\mathcal{PT}$-symmetric potentials $V(x,y)+iW(x,y)$ (i.e., $%
V(x,y)=V(-x,-y) $ and $W(-x,-y)=-W(x,y)$), which, like in the 1D setting,
admit particular exact solutions for $\alpha =1,2,$ and $3$, are given below
in Eqs. (\ref{nlm2D1}), (\ref{nlm2D2}), and (\ref{nlm2D3}), respectively,
along with the exact solutions (see Methods).

In particular, similar to 1D case, it is possible to find stationary
constant-intensity (CW) solutions of Eq.~(\ref{2D-NLS})
\begin{equation}
\psi (x,y,z)=Be^{i\theta (x,y)+igA^{2}z},\qquad B=\mathrm{const}  \label{ci2}
\end{equation}%
(cf. Eq. (\ref{C})) for a family of 2D complex potentials similar to their
1D counterparts (\ref{v}):
\begin{equation*}
V(x,y)=-\nabla \theta \lbrack (\nabla \theta )\mathbf{m}],\quad
W(x,y)=\nabla \lbrack (\nabla \theta )\mathbf{m}],
\end{equation*}%
where $\theta (x,y)$ is an arbitrary real function. 2D CW solutions (\ref%
{ci2}) will be considered in detail elsewhere.

%%%%%%%%%%%%%%%%%%%%%%%%%%%%%%%%%%%%%%%%%%%%%%%%%%%%%
\begin{figure}[tbp]
\begin{center}
{\scalebox{0.46}[0.4]{\includegraphics{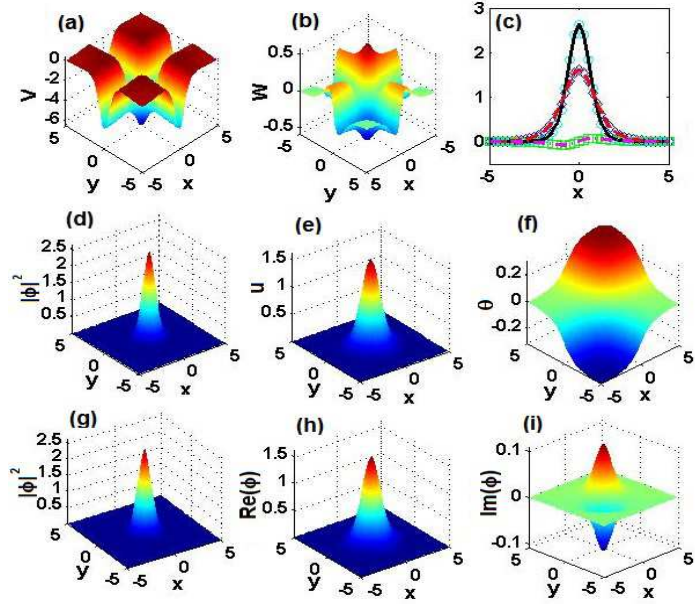}}}
\end{center}
\par
\vspace{-0.25in}
\caption{{\protect\small \textbf{2D $\mathcal{PT}$-symmetric solitons for $%
\protect\alpha =1$}. (a) Real and (b) imaginary parts of the 2D $\mathcal{PT}
$-symmetric complex potentials given by Eq. (\protect\ref{nlm2D1}). (c)
Comparison of real parts, imaginary parts, and intensity of exact soliton
solution $\protect\phi _{e}(x,0)$ and numerically found fundamental soliton $%
\protect\phi _{n}(x,0)$ at $\protect\mu =9/2$ , cf. Fig.~\protect\ref{num1}%
(a1) pertaining to the 1D setting. (d) The intensity, (e) amplitude, and (f)
phase of the exact soliton. (g) The intensity, (h) real part, and (i)
imaginary part of the numerically found fundamental soliton. Parameters are $%
(m_{0},\protect\theta _{0})=(0.5,0.1)$. }}
\label{num2D1}
\end{figure}
%%%%%%%%%%%%%%%%%%%%%%%%%%%%%%%%%%%%%%%%%%%%%%%%%%%%

%%%%%%%%%%%%%%%%%%%%%%%%%%%%%%%%%%%%%%%%%%%%%%%%%%%%%
\begin{figure}[tbp]
\begin{center}
{\scalebox{0.5}[0.4]{\includegraphics{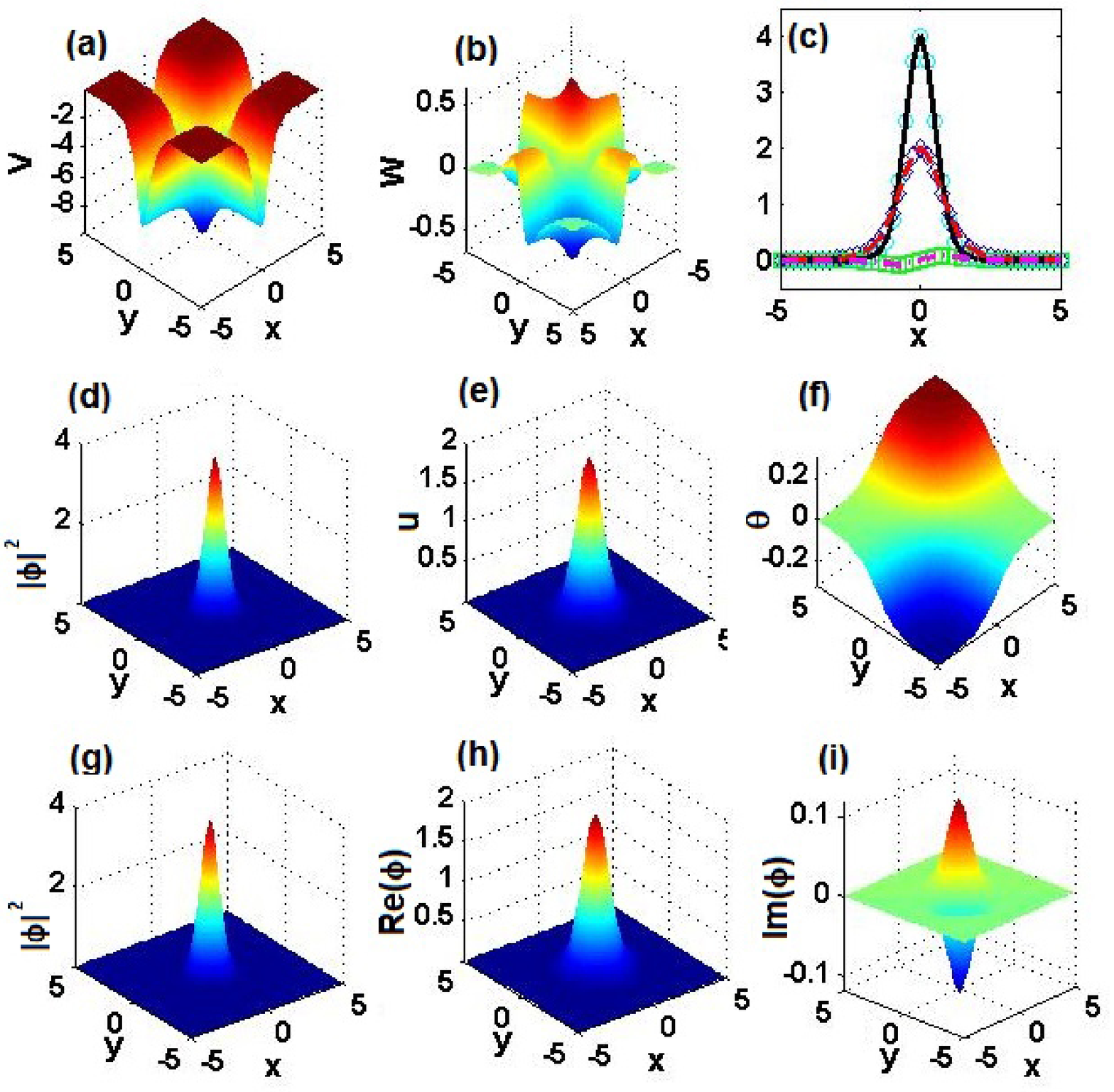}}}
\end{center}
\par
\vspace{-0.25in}
\caption{{\protect\small \textbf{2D $\mathcal{PT}$-symmetric nonlinear waves
for $\protect\alpha =2$}. The same as in Fig. \protect\ref{num2D1}, but for }%
$\protect\alpha =2$, and $\protect\mu =8$ in panel (c).{\protect\small \
Parameters are $(m_{0},\protect\theta _{0})=(0.4,0.1)$. }}
\label{num2D2}
\end{figure}
%%%%%%%%%%%%%%%%%%%%%%%%%%%%%%%%%%%%%%%%%%%%%%%%%%%%

%%%%%%%%%%%%%%%%%%%%%%%%%%%%%%%%%%%%%%%%%%%%%%%%%%%%%
\begin{figure}[tbp]
\begin{center}
{\scalebox{0.5}[0.4]{\includegraphics{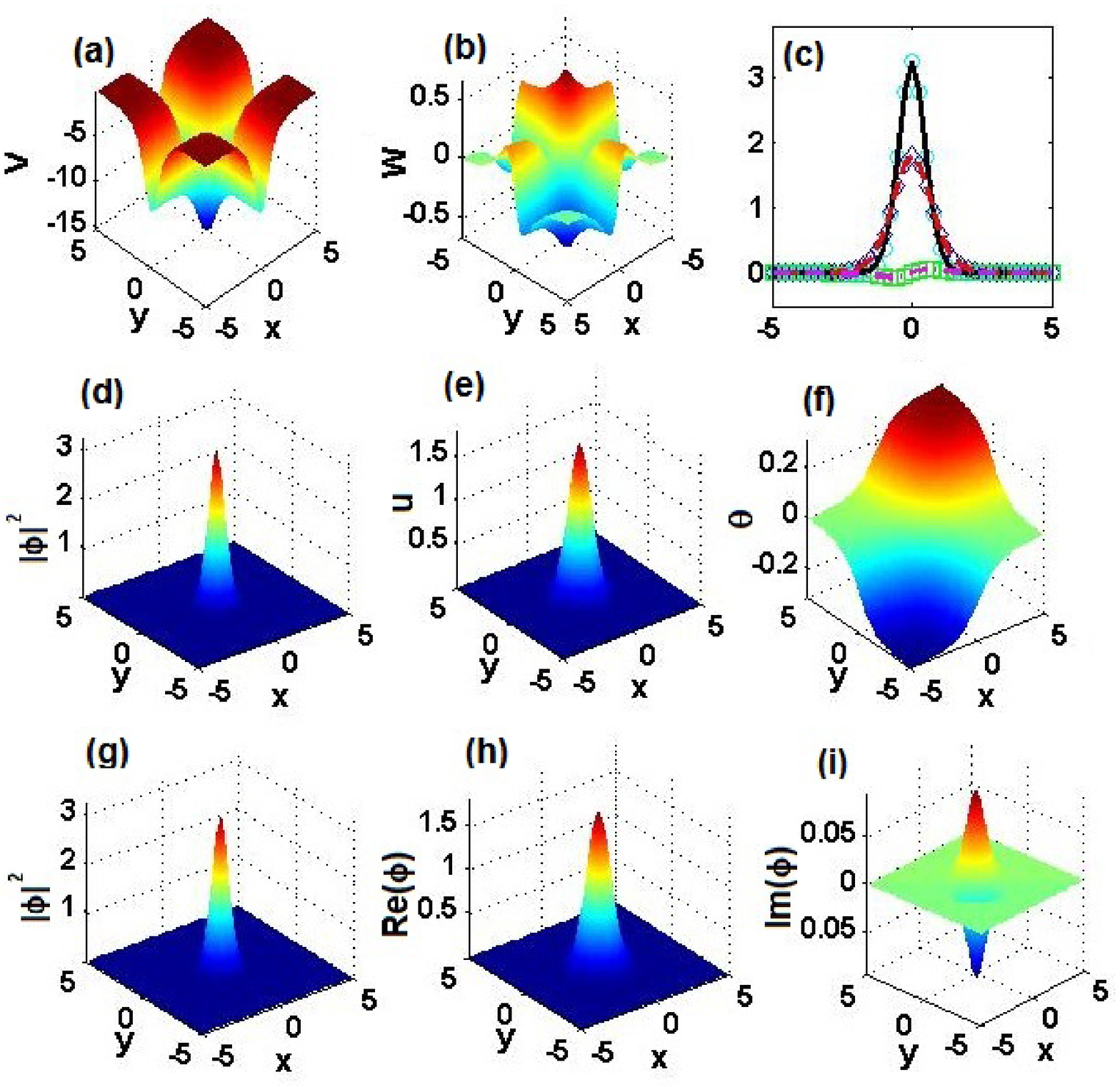}}}
\end{center}
\par
\vspace{-0.25in}
\caption{{\protect\small \textbf{2D $\mathcal{PT}$-symmetric nonlinear waves
for $\protect\alpha =3$}. The same as in Fig. \protect\ref{num2D1}, but for }%
$\protect\alpha =3$, and $\protect\mu =25/2$ in panel (c). {\protect\small \
Parameters are $(m_{0},\protect\theta _{0})=(0.2,0.1)$. }}
\label{num2D3}
\end{figure}
%%%%%%%%%%%%%%%%%%%%%%%%%%%%%%%%%%%%%%%%%%%%%%%%%%%%

%%%%%%%%%%%%%%%%%%%%%%%%%%%%%%%%%%%%%%%%%%%%%%%%%%%%%
\begin{figure}[tbp]
\begin{center}
{\scalebox{0.56}[0.56]{\includegraphics{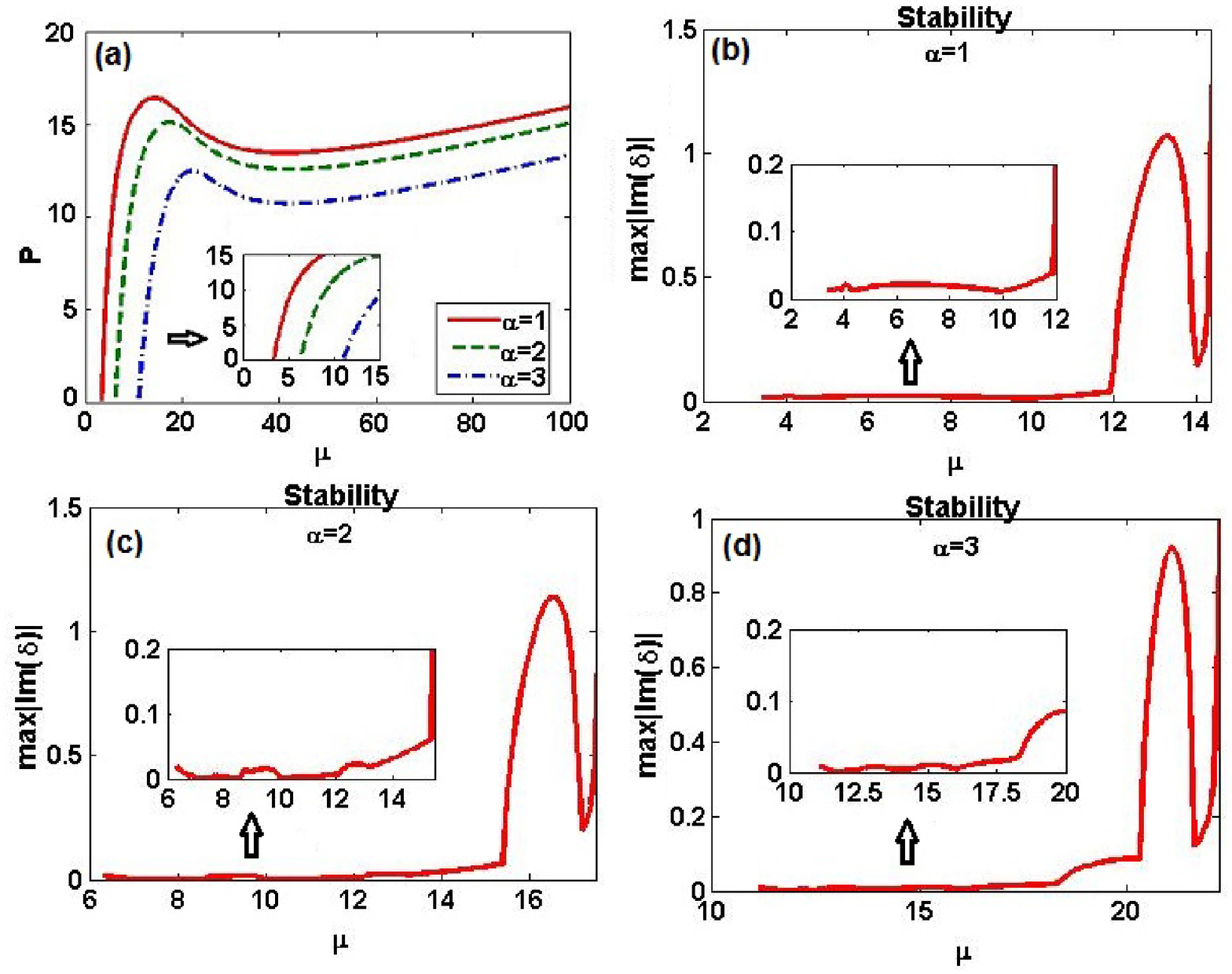}}}
\end{center}
\par
\vspace{-0.25in}
\caption{{\protect\small \textbf{2D integral power $P$ and linear stability
versus the soliton's propagation constant, $-\protect\mu $}. (a) Power $P$
versus $\protect\mu $. For the three different cases, $(\protect\alpha %
,m_{0},\protect\theta _{0})=(1,0.5,0.1),(2,0.4,0.1),(3,0.2,0.1)$, there are
no upper cutoffs of the existence domains for the solitons. The
corresponding lower cutoffs are $\protect\mu _{1}\approx 3.4,\protect\mu %
_{2}\approx 6.3,\protect\mu _{3}\approx 11.1$, respectively. (b, c, d): The
linear-stability spectra of numerically found fundamental solitons versus $%
\protect\mu $, around the corresponding exact nonlinear modes at $\protect%
\mu =9/2,8,25/2$, respectively. The insets in (b, c, d) clearly show a
portion of stability results when $\protect\mu$ is located in lower and
smaller intervals. }}
\label{pvsmu}
\end{figure}
%%%%%%%%%%%%%%%%%%%%%%%%%%%%%%%%%%%%%%%%%%%%%%%%%%%%

\vspace{0.1in} \noindent \textbf{Comparison of exact 2D solitons and
numerical solutions.} For $\alpha =1,2,3$, we first choose three parameter
sets, $(m_{0},\theta _{0})=(0.5,0.1),(0.4,0.1),(0.2,0.1)$, with the
corresponding 2D $\mathcal{PT}$-symmetric complex potentials shown in Figs.~%
\ref{num2D1}(a,b), \ref{num2D2}(a,b), and \ref{num2D3}(a,b), respectively.
The corresponding exact soliton solutions, given by Eqs.~(\ref{nlm2D1}), (%
\ref{nlm2D2}), and (\ref{nlm2D3}), with $\mu =9/2,8,$ and $25/2$, are shown
in the second row of Figs. \ref{num2D1}, \ref{num2D2}, \ref{num2D3},
respectively.

To verify the analytical soliton solutions, we have numerically found the
corresponding stationary fundamental solitons of Eq.~(\ref{2D-NLS}), which
are displayed in the third row of Figs.~\ref{num2D1}, \ref{num2D2}, \ref%
{num2D3}, respectively. As is shown in Figs.~\ref{num2D1}(c), \ref{num2D2}%
(c), and \ref{num2D3}(c), the difference between the exact solutions and
their numerical counterparts falls below $10^{-9}$, hence the exact
solutions are correct.

Furthermore, using the numerical method, we calculate the integral power at
other values of the soliton parameter $\mu $ and identify the existence
ranges of the numerically found solitons, as shown in Fig.~\ref{pvsmu}(a),
where the lower cutoffs are $\mu _{1}\approx 3.4,\mu _{2}\approx 6.3,\mu
_{3}\approx 11.1$, and there are no finite upper cutoffs. These numerical
soliton solutions are more often than not unstable, especially for larger
values of $\mu $. We display their linear-stability spectra in Figs.~\ref%
{pvsmu}(b,c,d) around $\mu =9/2,8,$ and $25/2$, respectively, i.e., around
the values at which the corresponding exact nonlinear modes exist. As a
result, we find that in first case, $(\alpha ,m_{0},\theta _{0})=(1,0.5,0.1)$%
, almost all the corresponding numerical and exact solitons are unstable;
however, in the second and third cases, \textit{viz}., $(\alpha
,m_{0},\theta _{0})=(2,0.4,0.1)$ and $(3,0.2,0.1)$, there exist some
stability regions, which are distributed around $\mu =8$ and $25/2$. To
confirm these linear-stability results, in what follows we investigate the
dynamics by dint of direct numerical simulations.

%%%%%%%%%%%%%%%%%%%%%%%%%%%%%%%%%%%%%%%%%%%%%%%%%%%%%
\begin{figure}[tbp]
\begin{center}
\vspace{0.05in} \hspace{-0.05in}{\scalebox{0.6}[0.56]{%
\includegraphics{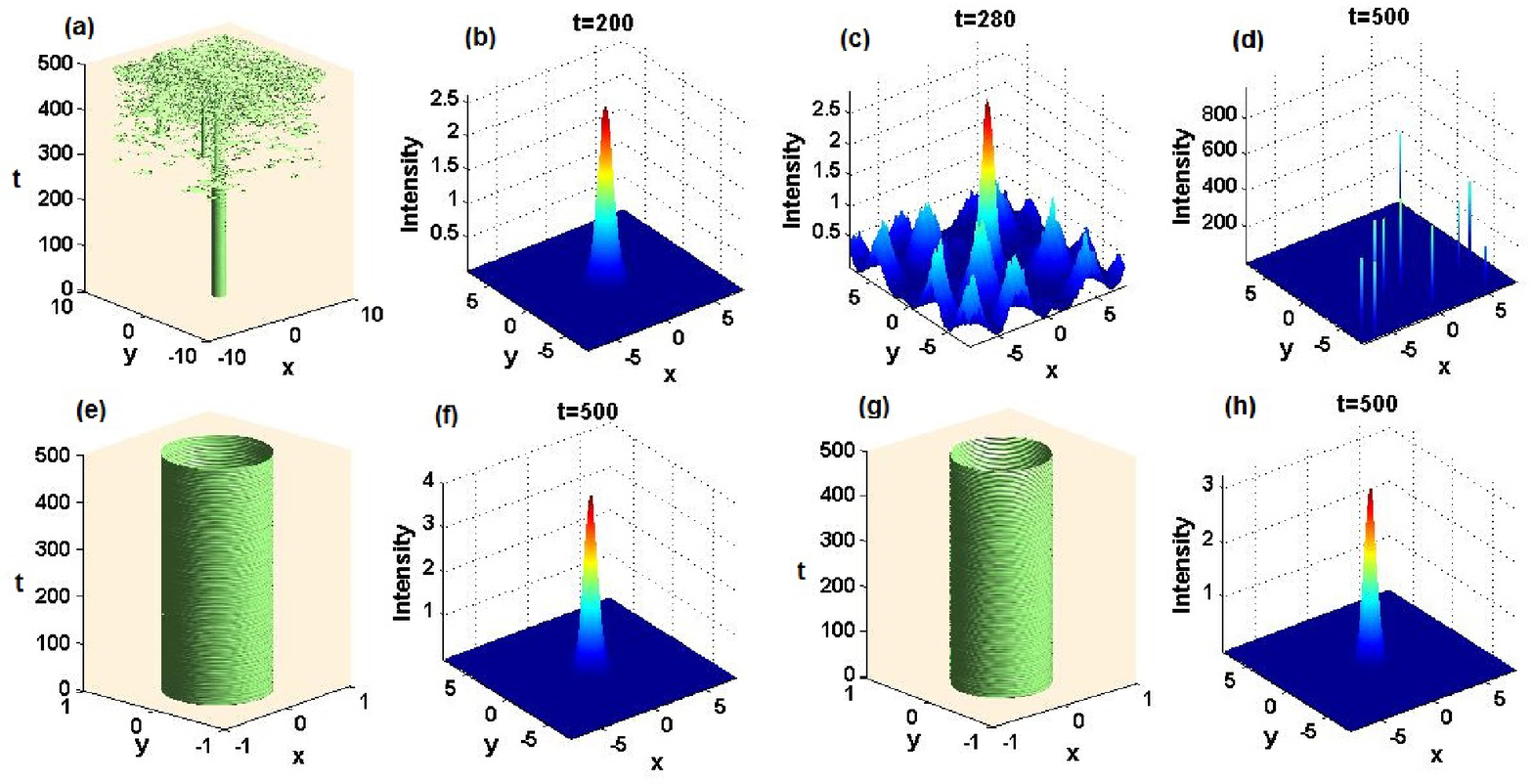}}}
\end{center}
\par
\vspace{-0.25in}
\caption{{\protect\small \textbf{Stable and unstable evolution of 2D
solitons.} (a) The unstable evolution in terms of the intensity isosurfaces,
corresponding to half the maximum initial intensity, of the 2D soliton from
Fig.~\protect\ref{num2D1} (these and other results are virtually identical
for exact solitons and their numerically found counterparts); (b) apparently
stable and (c) becoming-unstable intermediate states; (d) the unstable final
state. (e) The stable evolution of the soliton from Fig.~\protect\ref{num2D2}%
, and (f) its final shape. (g) The stable evolution of the soliton from Fig.~%
\protect\ref{num2D3}, and (h) its final shape. }}
\label{exa2}
\end{figure}
%%%%%%%%%%%%%%%%%%%%%%%%%%%%%%%%%%%%%%%%%%%%%%%%%%%%

%%%%%%%%%%%%%%%%%%%%%%%%%%%%%%%%%%%%%%%%%%%%%%%%%%%%%
\begin{figure}[tbp]
\begin{center}
{\scalebox{0.7}[0.6]{\includegraphics{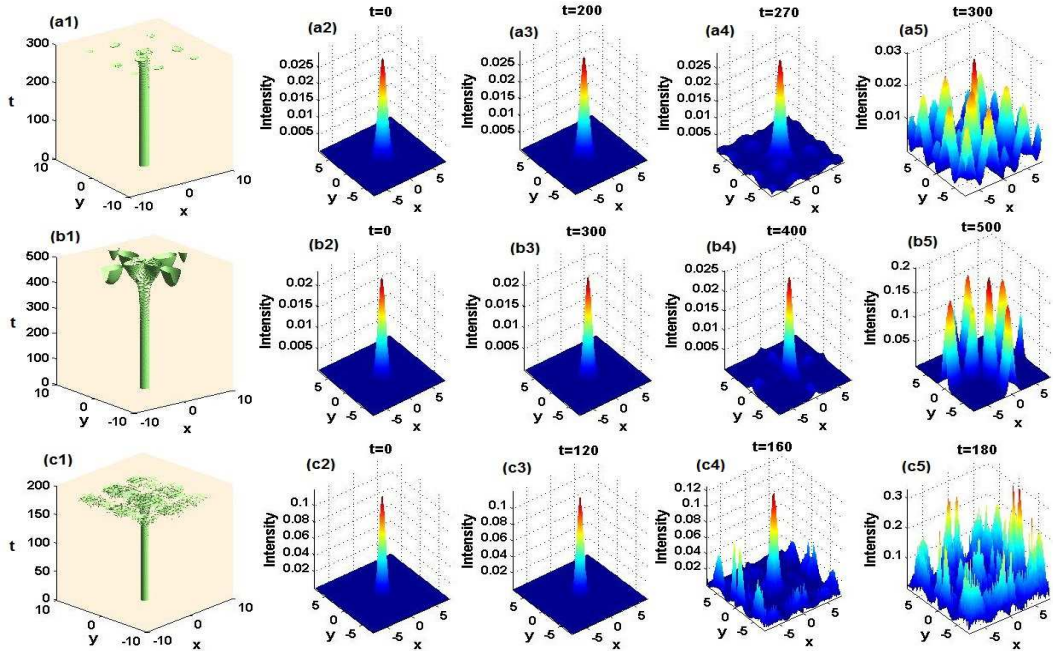}}}
\end{center}
\par
\vspace{-0.25in}
\caption{{\protect\small \textbf{Unstable evolutions of 2D solitons.}
(a1,b1,c1) The unstable evolution of numerical found solitons corresponding
to }${\protect\mu }${$_{1}=3.4,\protect\mu _{2}=6.3,\protect\mu _{3}=11.1$
for $\protect\alpha =1,2,3$, respectively. (a2, b2, c2) The corresponding
initial states; (a3,b3,c3) apparently stable intermediate states; (a4,b4,c4)
becoming-unstable intermediate states; (a5,b5,c5) unstable final states.}}
\label{num2}
\end{figure}
%%%%%%%%%%%%%%%%%%%%%%%%%%%%%%%%%%%%%%%%%%%%%%%%%%%%

\vspace{0.1in} \noindent\textbf{Dynamical behavior of 2D nonlinear modes.}
To display the evolution of the 2D exact or numerically found nonlinear
modes, we plot the corresponding intensity isosurfaces. For $\alpha =1$, as
shown in Figs.~\ref{exa2}(a,d), instability is produced by simulations of
the long-time evolution, although at shorter times, such as $t=200$, the
solution seems as a stable one, see Fig.~\ref{exa2}(b). The instability sets
in at $t\approx 280$ (Fig.~\ref{exa2}(c)). However, for $\alpha =2$ and $3$,
both Figs.~\ref{exa2}(e) and \ref{exa2}(g) exhibit fully robust evolution of
the initial-state solitons taken from Figs.~\ref{num2D2} and \ref{num2D3},
respectively. To confirm their stability, we display the corresponding
final-state soliton profiles in Figs.~\ref{exa2}(f) and \ref{exa2}(h), which
are identical to the corresponding initial profiles in Fig.~\ref{num2D2} and
Fig.~\ref{num2D3}.

We have also numerically investigated the stability of the solitons at the
lowest-power points $\mu _{1}=3.4,\mu _{2}=6.3,$ and $\mu _{3}=11.1$ for $%
\alpha =1,2,$ and $3$, respectively. They all turn out to be unstable,
although they may seem stable at a shorter propagation distance. Figure~\ref%
{num2} specifically exhibits their evolution and the corresponding
intermediate and final states. These results indicate a somewhat surprising
fact that the localized nonlinear mode with the lowest power is not
necessarily stable. On the other hand systematic simulations demonstrate
that the numerically found solitons are fully stable in the vicinity of the
corresponding exact nonlinear modes, that is, near $\mu _{1}=9/2,\mu
_{2}=8,\mu _{3}=25/2$ for $\alpha =1,2,3$. For this reason, the exact
soliton solutions are especially important ones, as they help to spot
stability areas for broad soliton families.

It is worthy to note that the dynamical-stability results are in good
agreement with the predictions produced above by the linear-stability
analysis (see Figs.~\ref{pvsmu}(b,c,d)). Thus, the latter analysis, in the
combination with systematic direct simulations, make it possible to identify
soliton-stability regions in a reliable form.

\vspace{0.1in} \noindent{\large \textbf{Discussion}}

We have reported analytical and numerical results for new classes of 1D and
2D stable spatial solitons in cubic nonlinear media with the $\mathcal{PT}$%
-symmetric generalized Scarf-II potentials and variable (position dependent)
diffraction coefficients. First, in the linear version of the model, linear
problem, parameter regions of the unbroken and broken $\mathcal{PT}$
symmetry have been numerically delineated. Then, in the presence of the Kerr
nonlinearity, particular exact solutions for nonlinear localized modes with
real eigenvalues have been obtained in the analytical form, and verified
numerically. These solitons are shown to be stable through the
linear-stability analysis and by means of direct simulations, in wide ranges
of the governing parameters. It is worthy to note that the addition of the
Kerr nonlinearity can \emph{fix} the broken $\mathcal{PT}$ symmetry of the
linear system, transforming complex eigenvalues into real ones. In addition,
stable bright solitons have been found in parameter regions where the $%
\mathcal{PT}$ symmetry of the linear states is broken, for various shapes of
the underlying real part of the potential, such as single- and double-well
forms. Finally, interactions and adiabatic transformations of the exact
solitons have been studied in detail, and the existence range and
propagation dynamics of numerically found solitons have been examined too.
We also study the evolution of constant-intensity waves in a model combining
the variable diffraction coefficient and complex potentials with globally
balanced gain and loss, which are more general than $\mathcal{PT}$-symmetric
ones, but feature similar properties. These theoretical results suggest new
experiments for $\mathcal{PT}$-symmetric nonlinear waves in nonlinear and
nonuniform optical media, and provide useful theoretical guidance for
studies in related fields, such as BECs.

\vspace{0.1in} \noindent{\large \textbf{Methods}}

\noindent \textbf{Nonlinear stationary modes.} Stationary solution of Eq.~(%
\ref{nls}) are looked for in the usual form, $\psi (x,z)=\phi (x)e^{i\mu z}$%
, where $-\mu $ is the propagation constant, and the localized complex wave
function satisfies the following ordinary differential equation with the $x$%
-dependent diffraction coefficient:
\begin{equation}
\left[ \frac{d}{dx}m(x)\frac{d}{dx}-V(x)-iW(x)+g|\phi |^{2}\right] \phi =\mu
\phi .
\end{equation}%
We take the complex wave function as $\phi (x)=u(x)\exp \left(
i\int_{0}^{x}v(s)ds\right) $ with real amplitude $u(x)$, and superfluid
phase velocity given by
\begin{equation}
v(x)=\frac{1}{m(x)u^{2}(x)}\int_{0}^{x}W(s)u^{2}(s)ds.
\end{equation}%
The amplitude satisfies the following second-order ordinary differential
equation:
\begin{equation}
\frac{d}{dx}\left[ m(x)\frac{du}{dx}\right] =[V(x)+m(x)v^{2}(x)-gu^{2}+\mu
]u,  \label{stat}
\end{equation}%
which may be transformed by setting $\tilde{u}(x)\equiv m(x)u_{x}$ into a
system of coupled first-order equations:
\begin{eqnarray}
&&\frac{du(x)}{dx}=\frac{\tilde{u}(x)}{m(x)},  \label{ode1} \\
&&\frac{d\tilde{u}(x)}{dx}=[V(x)+m(x)v^{2}(x)-gu^{2}+\mu ]u,  \label{ode2}
\end{eqnarray}%
solvable as a boundary-value problem by means of standard shooting methods
\cite{em3}. To achieve a higher precision and computation speed, we actually
used the spectral renormalization method \cite{spre} with some necessary
modifications. The method is spectrally efficient and relatively easy to
implement not only in the 1D case but also in the higher-dimensional
settings.

\vspace{0.1in}\noindent \textbf{\ Implementation of the numerical solution.}
To construct 1D localized solutions, one first needs to develop a convergent
iteration, to guarantee that the amplitude neither blows up nor decays to
zero. This may be realized by setting $\phi (x)=\lambda w(x)$, where $%
\lambda $ is a constant to be determined. Using the Fourier transform and
the modified spectral renormalization method \cite{spre}, we thus arrive as
the following iteration scheme:
\begin{equation}
\hat{w}=\frac{1}{k_{x}^{2}+p}F_{1}+\frac{\lambda ^{2}}{k_{x}^{2}+p}F_{2},
\label{ite1D}
\end{equation}%
where $F_{1}=\mathcal{F}\{[m_{x}\mathcal{F}^{-1}(ik_{x}\hat{w})-(V+iW)w-\mu
w+pmw]/m\}$, $F_{2}=\mathcal{F}(g|w|^{2}w/m),\lambda ^{2}=\int_{-\infty
}^{+\infty }[(k_{x}^{2}+p)|\hat{w}|^{2}-F_{1}\hat{w}^{\ast }]\mathrm{d}%
k_{x}/\int_{-\infty }^{+\infty }F_{2}\hat{w}^{\ast }\mathrm{d}k_{x}$, $%
\mathcal{F}$ denotes the 1D Fourier transform, and $p$ is an appropriate
positive constant (here $p=10$ is taken for the 1D case). A Gaussian or sech
can be taken as an input, which eventually leads to absolute errors $<10^{-9}
$, in both the convergence criterion and the numerically obtained solution
satisfying Eq.~(\ref{stat}). In general, one may restrict the number of
iterations to $<100$; however, for the sake of high precision, we admitted
up to $1000$ iterative steps. Once the above-mentioned conditions for two
absolute errors are satisfied simultaneously, the desired numerical soliton
is obtained as $\phi (x)=\lambda \mathcal{F}^{-1}[\hat{w}(x)]$.

In the 2D case, we needed to accordingly change $F_{1}$ in Eq.~(\ref{ite1D}%
), although the iteration scheme ran similar to its counterpart in the 1D
case:
\begin{equation}
\hat{w}=\frac{1}{k_{x}^{2}+k_{y}^{2}+p}F_{1}+\frac{\lambda ^{2}}{%
k_{x}^{2}+k_{y}^{2}+p}F_{2},  \label{ite2D}
\end{equation}%
where $F_{1}=\mathcal{F}\{[m(x)-1]\mathcal{F}^{-1}(-k_{x}^{2}\hat{w}%
)+[m(y)-1] \mathcal{F}^{-1}(-k_{y}^{2}\hat{w})+m_{x}(x)\mathcal{F}^{-1}(ik_x%
\hat{w})+ m_{y}(y)\mathcal{F}^{-1}(ik_y\hat{w})-(V+iW)w-\mu w+pw\}$, $%
\mathcal{F} $ denotes, the 2D Fourier transformation, and $p$ is an
appropriate positive constant (here $p=100$ is taken for the 2D case). Other
settings and procedures are similar to those in the 1D case.

\vspace{0.1in} \noindent \textbf{The linear-stability analysis.} For the
given position-dependent function $m(x)$ and complex-valued $\mathcal{PT}$%
-symmetric potential $V(x)+iW(x)$, one may solve Eq.~(\ref{stat}) (or
equivalently, Eqs.~(\ref{ode1}) and (\ref{ode2})), to obtain stationary
soliton solutions $\phi (x)$, by analytical or the above-mentioned methods.
Then localized nonlinear modes of Eq.~(\ref{nls}) can be found in the
stationary form, as $\psi (x,z)=\phi (x)e^{i\mu z}$. To explore the linear
stability of the localized modes in the 1D case, we consider a perturbed
solution~\cite{stable1, stable2},
\begin{equation}
\psi (x,z)=\{\phi (x)+\epsilon \lbrack F(x)e^{i\delta z}+G^{\ast
}(x)e^{-i\delta ^{\ast }z}]\}e^{i\mu z},  \label{pert}
\end{equation}%
where $\epsilon $ is an infinitesimal perturbation amplitude, $F(x)$ and $%
G(x)$ are the eigenfunctions of the linearized problem, and $-\delta $ is
the respective eigenfrequency, the instability taking place if some
eigenvalues are not purely real. Inserting the perturbed solution (\ref{pert}%
) into Eq.~(\ref{nls}) and linearizing with respect to $\epsilon $, we
obtain the following eigenvalue problem:
\begin{equation}
\left(%
\begin{array}{cc}
\hat{L}_{1} & g\phi ^{2}\vspace{0.05in} \\
-g \phi^{\ast 2} & -\hat{L}_{1}^{\ast }%
\end{array}%
\right) \left(
\begin{array}{c}
F(x)\vspace{0.05in} \\
G(x) \\
\end{array}%
\right) =\delta \left(
\begin{array}{c}
F(x)\vspace{0.05in} \\
G(x) \\
\end{array}%
\right) ,  \label{stable}
\end{equation}%
where $\hat{L}_{1}\equiv \partial _{x}\left( m(x)\partial _{x}\right)
-[V(x)+iW(x)]+2g|\phi |^{2}-\mu $. The $\mathcal{PT}$-symmetric nonlinear
modes are linearly stable provided $\delta $ has no imaginary part,
otherwise they are linearly unstable. The whole stability spectrum $\delta $
can be numerically calculated by the Fourier collocation method (see~\cite%
{stable2}).

For the 2D case (\ref{2D-NLS}), the operator $\hat{L}_{1}$ in Eq.~(\ref%
{stable}) is changed to $\hat{L}_{1}=\partial _{x}[m(x)\partial
_{x}]+\partial _{y}[m(y)\partial _{y}]-[V(x,y)+iW(x,y)]+2g|\phi |^{2}-\mu $.
Other technical details are similar to those in the 1D case. As concerns the
numerical computation of the full stability spectrum in 2D, the
Fourier-collocation method is usually of low precision for a small number of
Fourier modes. If one increases the number of the modes for a higher
accuracy, the necessary size of the dense matrix corresponding to the
eigenvalue problem may become prohibitively large \cite{stable2}. Therefore
the full 2D linear-stability spectrum in the $(m_0, \theta_0)$ space is not
displayed any more, because of the necessary large space size and number of
Fourier modes. But we will roughly depict the 2D linear-stability spectrum
with respect to the soliton parameter $\mu$ (see Figs.~\ref{pvsmu}(b, c,
d)), in contrast to the dynamical stability of long-time wave propagation.
By the way, on the account of the same reason that large spatial domains and
number of the Fourier modes are necessary for a high accuracy in 2D, the
corresponding $\mathcal{PT}$-symmetric linear spectra are not exhibited
further. Nevertheless, through repeated numerical tests, we find that it is
instructive that the $\mathcal{PT}$-symmetric breaking curves in 1D can
provide powerful reference for those in 2D.

\vspace{0.1in} \noindent \textbf{2D nonlinear modes for the $\mathcal{PT}$%
-symmetric potentials with different parameters $\alpha $.} The 2D $\mathcal{%
PT}$-symmetric potentials and exact solutions of Eqs. (\ref{2dsys1}) and (%
\ref{2dsys2}), which they admit, are listed as follows:

\textit{Case 1 ($\alpha =1$)}:
\begin{subequations}
\begin{eqnarray}
&&\mathbf{m}_{1}(x,y)=\mathrm{diag}\,(m_{0}\mathrm{sech}x+1,m_{0}\mathrm{sech%
}y+1),  \notag \\
&&V_{1}(x,y)\!=\!\frac{1}{4}\!\!\sum_{\sigma =x,y}\!\!\left[ 15m_{0}\mathrm{%
sech}\sigma -\!(4\theta _{0}^{2}\!+\!15)\mathrm{sech}^{2}\sigma
-m_{0}(4\theta _{0}^{2}+21)\mathrm{sech}^{3}\sigma \right] +m_{0}(\theta
_{0}^{2}+\frac{21}{4})(\mathrm{sech}x\,\mathrm{sech}y)^{3},\qquad  \notag \\
&&W_{1}(x,y)=-\theta _{0}\sum_{\sigma =x,y}\mathrm{sech}\sigma \,\mathrm{tanh%
}\sigma (5m_{0}\mathrm{sech}\sigma +4),  \notag \\
&&\phi _{1}(x,y)=\sqrt{m_{0}(4\theta _{0}^{2}+21)/(4g)}\,(\mathrm{sech}x\,%
\mathrm{sech}y)^{3/2}\exp \left\{ i\theta _{0}[\mathrm{tan^{-1}}(\mathrm{sinh%
}x)+\mathrm{tan^{-1}}(\mathrm{sinh}y)]\right\} ,\qquad  \label{nlm2D1}
\end{eqnarray}%
where $\mu =9/2$.

\textit{Case 2 ($\alpha =2$):}
\end{subequations}
\begin{subequations}
\begin{eqnarray}
&&\mathbf{m}_{2}(x,y)=\mathrm{diag}\,(m_{0}\mathrm{sech}^{2}x+1,m_{0}\mathrm{%
sech}^{2}y+1),  \notag \\
&&V_{2}(x,y)\!=\!\sum_{\sigma =x,y}\left[ (8m_{0}-\theta _{0}^{2}-6)\mathrm{%
sech}^{2}\sigma -m_{0}(\theta _{0}^{2}+10)\mathrm{sech}^{4}\sigma \right]
+m_{0}(\theta _{0}^{2}+10)(\mathrm{sech}x\,\mathrm{sech}y)^{4},  \notag \\
&&W_{2}(x,y)=-\theta _{0}\sum_{\sigma =x,y}\mathrm{sech}\sigma \,\mathrm{tanh%
}\sigma (7m_{0}\mathrm{sech}^{2}\sigma +5),  \notag \\
&&\phi _{2}(x,y)=\sqrt{m_{0}(\theta _{0}^{2}+10)/g}\,(\mathrm{sech}x\,%
\mathrm{sech}y)^2\exp \left[i\theta _{0}\sum_{\sigma =x,y}\mathrm{tan^{-1}}(%
\mathrm{sinh}\sigma)\right],\qquad  \label{nlm2D2}
\end{eqnarray}%
where $\mu =8$.

\textit{Case 3 ($\alpha =3$):}
\end{subequations}
\begin{subequations}
\begin{eqnarray}
&&\mathbf{m}_{3}(x,y)=\mathrm{diag}\,(m_{0}\mathrm{sech}^{3}x+1,m_{0}\mathrm{%
sech}^{3}y+1),  \notag \\
&&V_{3}(x,y)\!=\!-\frac{1}{4}\!\!\sum_{\sigma =x,y}\!\!\left[ (4\theta
_{0}^{2}\!+\!35)\mathrm{sech}^{2}\sigma -55m_{0}\mathrm{sech}^{3}\sigma
+m_{0}(4\theta _{0}^{2}+65)\mathrm{sech}^{5}\sigma \right]  \notag \\
&&\qquad \qquad \qquad \qquad \qquad \qquad +\frac{m_{0}}{4}(4\theta
_{0}^{2}+65)(\mathrm{sech}x\,\mathrm{sech}y)^{5},\qquad  \notag \\
&&W_{3}(x,y)=-3\theta _{0}\sum_{\sigma =x,y}\mathrm{sech}\sigma \,\mathrm{%
tanh}\sigma (3m_{0}\mathrm{sech}^{3}\sigma +2),  \notag \\
&&\phi _{3}(x,y)=\sqrt{m_{0}(4\theta _{0}^{2}+65)/(4g)}\,(\mathrm{sech}x\,%
\mathrm{sech}y)^{5/2}\exp \left\{ i\theta _{0}[\mathrm{tan^{-1}}(\mathrm{sinh%
}x)+\mathrm{tan^{-1}}(\mathrm{sinh}y)]\right\} ,\qquad  \label{nlm2D3}
\end{eqnarray}%
where $\mu =25/2$.

\vspace{0.1in}

\noindent {\large \textbf{Acknowledgments}}\newline
{\small The authors thank the three referees for their valuable comments and suggestions. This work of Z.Y. and Y.C. was supported by the NSFC under Grant
No.11571346 and the Youth Innovation Promotion Association CAS. The work of
B.A.M. is supported, in part, by grant No. 2015616 from the joint program in
physics between NSF and Binational (US-Israel) Science Foundation.}

\vspace{0.15in}

\noindent{\large \textbf{Author contributions}} \newline
Z.Y. and Y.C. conceived the idea and presented the overall theoretical
analysis. Y.C. and Z.Y. discussed and performed the numerical experiments.
Y.C., Z.Y., D.M. and B.A.M. contributed to writing of the manuscript and in
discussing and interpreting the obtained results.

\vspace{0.15in}

\noindent{\large \textbf{Additional information}}\newline
Competing financial interests: The authors declare no competing financial
interests.
\end{subequations}


\begin{thebibliography}{99}
\bibitem{qm} {\small Barton, G. \textit{Introduction to Advanced Field
Theory} (Wiley, New York, 1963). }

\bibitem{bender1} {\small Bender, C. M. \& Boettcher, S. Real spectra in
non-Hermitian Hamiltonians having $\mathcal{PT}$ symmetry. \textit{Phys.
Rev. Lett.} \textbf{80}, 5243-5246 (1998). }

\bibitem{dorey} {\small Dorey, P., Dunning, C. \& Tateo, R. Spectral
equivalences, Bethe ansatz equations, and reality properties in $\mathcal{PT}
$-symmetric quantum mechanics. \textit{J. Phys. A: Math. Gen.} \textbf{34},
5679-5704 (2001). }

\bibitem{bender2} {\small Bender, C. M., Brody, D. C. \& Jones, H. F.
Complex extension of Quantum Mechanics. \textit{Phys. Rev. Lett.} \textbf{89}%
, 270401 (2002). }

\bibitem{bender3} {\small Bender, C. M. Making sense of non-Hermitian
Hamiltonians. \textit{Rep. Prog. Phys.} \textbf{70}, 947-1018 (2007). }

\bibitem{bender4} {\small Bender, C. M. Rigorous backbone of $\mathcal{PT}$%
-symmetric quantum mechanics. \textit{J. Phys. A: Math. Theor.} \textbf{49},
401002 (2016). }

\bibitem{review} {\small Makris, K. G., El-Ganainy, R., Christodoulides, D.
N. \& Musslimani Z. H. $\mathcal{PT}$\ symmetric periodic optical
potentials. \textit{Int. J. Theor. Phys.} \textbf{50}, 1019-1041 (2011).}

\bibitem{ptqm} {\small Moiseyev, N. \textit{Non-Hermitian Quantum Mechanics}
(Cambridge Univ. Press, 2011). }

\bibitem{unbreakable} {\small Kartashov, Y. V., Malomed, B. A. \& Torner, L.
Unbreakable $\mathcal{PT}$}{\small \ symmetry of solitons supported by
inhomogeneous defocusing nonlinearity. \textit{Opt. Lett.} \textbf{39},
5641-5644 (2014).}

\bibitem{theo1} {\small Ruschhaupt, A., Delgado, F. \& Muga, J. G. Physical
realization of $\mathcal{PT}$\ -symmetric potential scattering in a planar
slab waveguide. \textit{J. Phys. A: Math. Gen.} \textbf{38}, L171-L176
(2005).}

\bibitem{theo2} {\small El-Ganainy, R., Makris, K. G., Christodoulides, D.
N. \& Musslimani, Z. H. Theory of coupled optical $\mathcal{PT}$%
-symmetric structures. \textit{Opt. Lett.} \textbf{32}, 2632-2634 (2007). }

\bibitem{theo3} {\small Berry, M. V. Optical lattices with $\mathcal{PT}$%
-symmetry are not transparent. \textit{J. Phys. A: Math. Theor.} \textbf{41}%
, 244007 (2008). }

\bibitem{theo4} {\small Klaiman S., G\"{u}nther, U. \& Moiseyev, N.
Visualization of branch points in $\mathcal{PT}$-Symmetric Waveguides.
\textit{Phys. Rev. Lett.} \textbf{101}, 080402 (2008).}

\bibitem{theo5} {\small Longhi, S. Bloch oscillations in complex crystals
with $\mathcal{PT}$\ symmetry. \textit{Phys. Rev. Lett.} \textbf{103},
123601 (2009).}

\bibitem{exp1} {\small Guo, A. {\it et al.}
Observation of $\mathcal{PT}$-symmetry breaking in complex optical
potentials. \textit{Phys. Rev. Lett.} \textbf{103}, 093902 (2009).}

\bibitem{exp2} {\small R\"{u}ter, C. E. {\it et al.} Observation of parity-time
symmetry in optics. \textit{Nature Phys.} \textbf{6}, 192-195 (2010). }

\bibitem{exp3} {\small Regensburger, A. {\it et al.} Parity-time
synthetic photonic lattices. \textit{Nature} \textbf{488}, 167-171 (2012). }

\bibitem{exp4} {\small Castaldi, G., Savoia, S., Galdi, V., Al\`{u}, A. \&
Engheta, N. $\mathcal{PT}$ Metamaterials via Complex-Coordinate
Transformation Optics. \textit{Phys. Rev. Lett.} \textbf{110}, 173901
(2013). }

\bibitem{exp5} {\small Hodaei, H., Miri, M. A., Heinrich, M.,
Christodoulides, D. N. \& Khajavikhan, M. Parity-time-symmetric microring
lasers. \textit{Science} \textbf{346}, 975-978 (2014). }

\bibitem{exp6} {\small Peng, B., \"{O}zdemir, \c{S}. K., Chen, W., Nori, F.
\& Yang, L. Parity-time-symmetric whispering gallery microcavities. \textit{%
Nature Phys.} \textbf{10}, 394-398 (2014). }

\bibitem{exp7} {\small Wimmer, M. {\it et al.} Observation of optical solitons
in $\mathcal{PT}$-symmetric lattices. \textit{Nature Commun.} \textbf{6},
7782 (2015). }

\bibitem{exp8} {\small Zhang, Z. {\it et al.} Observation of
parity-time symmetry in optically induced atomic lattices. \textit{Phys.
Rev. Lett.} \textbf{117}, 123601 (2016). }

\bibitem{t1} {\small Musslimani, Z. H., Makris, K. G., El-Ganainy, R. \&
Christodoulides, D. N. Optical solitons in $\mathcal{PT}$ periodic
potentials. \textit{Phys. Rev. Lett.} \textbf{100}, 030402 (2008). }

\bibitem{shi2011} {\small Shi, Z., Jiang, X., Zhu, X. \& Li, H. Bright
spatial solitons in defocusing Kerr media with PT-symmetric potentials.
\textit{Phys. Rev. A} \textbf{84}, 053855 (2011).}

\bibitem{t2} {\small Makris, K. G., El-Ganainy, R., Christodoulides, D. N.
\& Musslimani, Z. H. Beam dynamics of $\mathcal{PT}$-symmetric optical
lattices. \textit{Phys. Rev. Lett}. \textbf{100}, 103904 (2008). }

\bibitem{Radik} {\small Driben, R. \& Malomed, B. A. Stability of solitons
in parity-time-symmetric couplers. \textit{Opt. Lett.} \textbf{36},
4323-4325 (2011).}

\bibitem{t3} {\small Abdullaev, F. Kh., Kartashov, Y. V., Konotop, V. V. \&
Zezyulin, D. A. Solitons in $\mathcal{PT}$-symmetric nonlinear lattices.
\textit{Phys. Rev. A} \textbf{83}, 041805 (2011). }

\bibitem{t4} {\small Li, K. \& Kevrekidis, P. G. $\mathcal{PT}$-symmetric
oligomers: Analytical solutions, linear stability, and nonlinear dynamics.
\textit{Phys. Rev. E} \textbf{83}, 066608 (2011). }

\bibitem{Barash} {\small Alexeeva, N. V., Barashenkov, I. V., Sukhorukov, A.
A. \& Kivshar, Y. S. Optical solitons in $\mathcal{PT}$-symmetric nonlinear
couplers with gain and loss. \textit{Phys. Rev. A} \textbf{85}, 063837
(2012).}

\bibitem{t5} {\small Zezyulin, D. A. \& Konotop, V. V. Nonlinear modes in
finite-dimensional $\mathcal{PT}$-symmetric systems. \textit{Phys. Rev. Lett.%
} \textbf{108}, 213906 (2012). }

\bibitem{t6} {\small Nixon, S., Ge, L. \& Yang, J. Stability analysis for
solitons in $\mathcal{PT}$-symmetric optical lattices. \textit{Phys. Rev. A}
\textbf{85}, 023822 (2012). }

\bibitem{t7} {\small Achilleos, V., Kevrekidis, P. G., Frantzeskakis, D. J.
\& Carretero-Gonz{\'a}lez, R. Dark solitons and vortices in $\mathcal{PT}$%
-symmetric nonlinear media: From spontaneous symmetry breaking to nonlinear $%
\mathcal{PT}$ phase transitions. \textit{Phys. Rev. A} \textbf{86}, 013808
(2012). }

\bibitem{t8} {\small Cartarius, H. \& Wunner, G. Model of a $\mathcal{PT}$%
-symmetric Bose-Einstein condensate in a $\delta$-function double-well
potential. \textit{Phys. Rev. A} \textbf{86}, 013612 (2012). }

\bibitem{t8a} {\small Yan, Z. Complex-symmetric nonlinear Schr\"odinger equation and Burgers equation.
\textit{ Phil. Trans. R. Soc. A} \textbf{371}, 20120059 (2013).}

\bibitem{t9} {\small Lumer, Y., Plotnik, Y., Rechtsman, M. C. \& Segev, M.
Nonlinearly induced $\mathcal{PT}$ transition in photonic systems. \textit{%
Phys. Rev. Lett.} \textbf{111}, 263901 (2013). }

\bibitem{Raymond} {\small Zhang, X. {\it et al.} Discrete solitons and scattering of lattice waves in guiding
arrays with a nonlinear }$\mathcal{PT}${\small -symmetric defect. \textit{%
Opt. Exp}. \textbf{22}, 13927-13939 (2014).}

\bibitem{Jennie} {\small D'Ambroise, J., Kevrekidis, P. G. \& Malomed, B. A.
Staggered parity-time-symmetric ladders with cubic nonlinearity. \textit{%
Phys. Rev. E} \textbf{91}, 033207 (2015).}

\bibitem{t10} {\small Yan, Z., Wen, Z. \& Konotop, V. V. Solitons in a
nonlinear Schr\"{o}dinger equation with $\mathcal{PT}$-symmetric potentials
and inhomogeneous nonlinearity: stability and excitation of nonlinear modes.
\textit{Phys. Rev. A} \textbf{92}, 023821 (2015). }

\bibitem{t11} {\small Makris, K. G., Musslimani, Z. H., Christodoulides, D.
N., \& Rotter, S. Constant-intensity waves and their modulation instability
in non-Hermitian potentials. \textit{Nature Commun.} \textbf{6}, 7257
(2015). }

\bibitem{t12} {\small Yan, Z., Wen, Z. \& Hang, C. Spatial solitons and
stability in self-focusing and defocusing Kerr nonlinear media with
generalized parity-time-symmetric Scarf-II potentials. \textit{Phys. Rev. E}
\textbf{92}, 022913 (2015). }

\bibitem{t13} {\small Wen, Z. \& Yan, Z. Dynamical behaviors of optical
solitons in parity-time ($\mathcal{PT}$) symmetric sextic anharmonic
double-well potentials. \textit{Phys. Lett. A} \textbf{379}, 2025-2029
(2015). }


\bibitem{t13a} {\small Wen, X. Y., Yan, Z. \& Yang, Y. Dynamics of higher-order rational solitons
for the nonlocal nonlinear Schr\"odinger equation with the self-induced parity-time-symmetric potential.
\textit{Chaos} \textbf{26}, 063123 (2016). }

\bibitem{t13b} {\small X. Li \& Yan, Z. Stability, integrability, and nonlinear dynamics of $\PT$-symmetric
optical couplers with cubic cross-interactions or cubic-quintic nonlinearities.
\textit{Chaos} \textbf{27}, 013105 (2017). }

\bibitem{t14} {\small Kartashov, Y. V., Konotop, V. V. \& Torner, L.
Topological states in partially-$\mathcal{PT}$-symmetric azimuthal
potentials. \textit{Phys. Rev. Lett.} \textbf{115}, 193902 (2015). }

\bibitem{t15} {\small Liu, B., Li, L. \& Mihalache, D. Vector soliton
solutions in $\mathcal{PT}$-symmetric coupled waveguides and their relevant
properties. \textit{Rom. Rep. Phys.} \textbf{67}, 802-818 (2015). }

\bibitem{t16} {\small He, Y., Zhu, X., Mihalache, D., Liu, J. \& Chen, Z.
Lattice solitons in $\mathcal{PT}$-symmetric mixed linear-nonlinear optical
lattices. \textit{Phys. Rev. A} \textbf{85}, 013831 (2012). }

\bibitem{t17} {\small Wang, H. {\it et al.} Two-dimensional solitons in triangular photonic
lattices with parity-time symmetry. \textit{Opt. Commun.} \textbf{335},
146-152 (2015). }

\bibitem{t17a} {\small He, Y., Zhu, X. \& Mihalache, D. Dynamics of spatial
solitons in parity-time-symmetric optical lattices: a selection of recent
theoretical results. \textit{Rom. J. Phys.} \textbf{61}, 595-613 (2016). }

\bibitem{t17b} {\small Li, P., Mihalache, D. \& Li, L. Asymmetric solitons
in parity-time-symmetric double-hump Scarf-II potentials. \textit{Rom. J.
Phys.} \textbf{61}, 1028-1039 (2016). }

\bibitem{t18} {\small Kartashov, Y. V., Hang, C., Huang, G. \& Torner, L.
Three-dimensional topological solitons in $\mathcal{PT}$-symmetric optical
lattices. \textit{Optica} \textbf{3}, 1048 (2016). }

\bibitem{t19} {\small Hahn, C. {\it et al.} Observation of exceptional points in reconfigurable
non-Hermitian vector-field holographic lattices. \textit{Nature Commun}.
\textbf{7}, 12201 (2016). }

\bibitem{t20} {\small Burlak, G., Garcia-Paredes, S. \& Malomed B. A. $%
\mathcal{PT}$-symmetric couplers with competing cubic-quintic
nonlinearities. \textit{Chaos} \textbf{26}, 113103 (2016).}

\bibitem{yansr16} {\small Chen, Y. \& Yan, Z. Solitonic dynamics and
excitations of the nonlinear Schr\"odinger equation with third-order
dispersion in non-Hermitian $\mathcal{PT}$-symmetric potentials. \textit{%
Sci. Rep.} \textbf{6}, 23478 (2016). }

\bibitem{yanchaos16} {\small Yan, Z., Chen, Y. \& Wen, Z. On stable
solitons and interactions of the generalized Gross-Pitaevskii equation with $%
\mathcal{PT}$- and non-$\mathcal{PT}$-symmetric potentials. \textit{Chaos}
\textbf{26}, 083109 (2016). }

\bibitem{yanstable2017} {\small Chen, Y. \& Yan, Z. Stable
parity-time-symmetric nonlinear modes and excitations in a derivative
nonlinear Schr\"{o}dinger equation. \textit{Phys. Rev. E} \textbf{95},
012205 (2017). }

\bibitem{pt-review-1} {\small Suchkov, S. V. {\it et al.} Nonlinear switching and solitons
in $\mathcal{PT}$-symmetric photonic systems, \textit{Laser Photonics Rev}.
\textbf{10}, 177 (2016).}

\bibitem{pt-review} {\small Konotop, V. V., Yang, J. \& Zezyulin, D. A.
Nonlinear waves in $\mathcal{PT}$-symmetric systems. \textit{Rev. Mod. Phys.}
\textbf{88}, 035002 (2016). }

\bibitem{em1} {\small Wannier, G. H. The structure of electronic excitation
levels in insulating crystals. \textit{Phys. Rev.} \textbf{52}, 191-197
(1937). }

\bibitem{em2} {\small Morrow, R. A. \& Brownstein, K. R. Model
effective-mass Hamiltonians for abrupt heterojunctions and the associated
wave-function-matching conditions. \textit{Phys. Rev. B} \textbf{30},
678-680 (1984). }

\bibitem{em3} {\small van Roos, O. Position-dependent effective masses in
semiconductor theory. \textit{Phys. Rev. B} \textbf{27}, 7547-7552 (1983). }

\bibitem{em4} {\small Morrow, R. A. Establishment of an effective-mass
Hamiltonian for abrupt heterojunctions. \textit{Phys. Rev. B} \textbf{35},
8074-8079 (1987). }

\bibitem{em5} {\small Paul, S. F. \& Fouckhardt, H. An improved shooting
approach for solving the time-independent Schr\"odinger equation for III/V
QW structures. \textit{Phys. Lett. A} \textbf{286}, 199-204 (2001). }

\bibitem{em45} {\small Konotop, V. V. On wave propagation in periodic
structures with smoothly varying parameters. \textit{J. Opt. Soc. Am. B}
\textbf{14}, 364-369 (1997). }

\bibitem{em6} {\small Midya, B., Roy B. \& Roychoudhury, R. Position
dependent mass Schr\"odinger equation and isospectral potentials:
Intertwining operator approach. \textit{J. Math. Phys.} \textbf{51}, 022109
(2010). }

\bibitem{em7} {\small F\"{o}rster, J., Saenz, A. \& Wolff U. Matrix
algorithm for solving Schr\"odinger equations with position-dependent mass
or complex optical potentials. \textit{Phys. Rev. E} \textbf{86}, 016701
(2012). }

\bibitem{em8} {\small Abdullaev, F. Kh. \& Garnier J. Solitons in media with
random dispersive perturbations. \textit{Physica D} \textbf{134}, 303-315
(1999).}

\bibitem{Inguscio} {\small Burger, S. {\it et al.} Superfluid and dissipative
dynamics of a Bose-Einstein condensate in a periodic optical potential.
\textit{Phys. Rev. Lett}. \textbf{86}, 4447-4450 (2001). }

\bibitem{Pitaevskii} {\small Kramer, M., Menotti, C., Pitaevskii, L. \&
Stringari, S. Bose-Einstein condensates in 1D optical lattices -
Compressibility, Bloch bands and elementary excitations. \textit{Eur. Phys. J%
}. D \textbf{27}, 247-261 (2003).}

\bibitem{Silberberg} {\small Eisenberg, H. S., Silberberg, Y., Morandotti,
R. \& Aitchison, J. S. Diffraction management. \textit{Phys. Rev. Lett}.
\textbf{85}, 1863-1866 (2000).}

\bibitem{Longhi} {\small Longhi, S. Quantum-optical analogies using photonic
structures. \textit{Laser Phot. Rev.} \textbf{3}, 243-261 (2009).}

\bibitem{Scarf} {\small Scarf, F. L. New soluble energy band problem.
\textit{Phys. Rev.} \textbf{112}, 1137-1140 (1958).}


\bibitem{BEC} {\small Brazhnyi, V. A. \& Konotop, V. V. Theory of nonlinear
matter waves in optical lattices. Mod. Phys. Lett. \textbf{18}, 627-651
(2004).}

%\bibitem{keer} {\small Agrawal, G. P. \textit{Nonlinear Fibre Optics} (5th
%ed.) (Academic Press, New York, 2014). }

%\bibitem{keer2} {\small Biswas, A., Milovic, D., \& Edwards, M. \textit{%
%Mathematical Therory of Dispersion-Managed Optical Solitons} (Higher
%Education Press, Beijing 2010). }

\bibitem{real41} {\small Bagchi, B. \& Quesne, C. sl(2,}$C${\small ) as a
complex Lie algebra and the associated non-Hermitian Hamiltonians with real
eigenvalues. \textit{Phys. Lett. A} \textbf{273}, 285-292 (2000). }

\bibitem{real42} {\small Bagchi, B., Quesne, C. \& Znojil M. Generalized
Continuity equation and modified normalization in $\mathcal{PT}$-symmetric
quantum mechanics. \textit{Mod. Phys. Lett. A} \textbf{16}, 2047-2057
(2001). }

\bibitem{real4} {\small Ahmed, A. Real and complex discrete eigenvalues in
an exactly solvable one-dimensional complex $\mathcal{PT}$-invariant
potential. \textit{Phys. Lett. A} \textbf{282}, 343-348 (2000). }

\bibitem{spm} {\small Trefethen, L. N. \textit{Spectral Methods in Matlab}
(SIAM, 2000). }

\bibitem{sphm} {\small Shen, J. \& Tang, T. \textit{Spectral and high-order
methods with applications} (Science Press, Beijing, 2006). }

\bibitem{Abdullaev} {\small Tsoy, E. N., Alaayarov, I. M. \& Abdullaev, F.
Kh. Stable localized modes in asymmetric waveguides with gain and loss.
\textit{Opt. Lett}. \textbf{39}, 4215-4218 (2014).}

\bibitem{Lisbon} {\small Konotop, V. V. \& Zezyulin, D. A. Families of
stationary modes in complex potentials. \textit{Opt. Lett.} \textbf{39},
5355-5538 (2014).}

\bibitem{Vermont} {\small Nixon, S. \& Yang, J. Bifurcation of soliton
families from linear modes in non-}$PT${\small -symmetric complex
potentials. \textit{Stud. Appl. Math}. \textbf{136}, 459-483 (2016).}

\bibitem{wadati} {\small Wadati, M. Construction of parity-time symmetric
potential through the soliton theory. \textit{J. Phys. Soc. Jpn.}. \textbf{77%
},  074005 (2008).}

\bibitem{opso} {\small Kivshar, Y. S. \& Agrawal, G. P. \textit{Optical
Solitons: From Fibers to Photonic Crystals} (Academic, San Diego, 2003). }

\bibitem{op2} {\small Malomed, B. A., Mihalache, D., Wise, F. \& Torner, L.
Spatiotemporal optical solitons. \textit{J. Opt. B: Quantum Semiclassical
Opt.} \textbf{7}, R53-R72 (2005). }

\bibitem{op2a} {\small Mihalache, D. Localized structures in nonlinear
optical media: a selection of recent studies. \textit{Rom. Rep. Phys.}
\textbf{67}, 1383-1400 (2015). }

\bibitem{op2b} {\small Malomed, B. A., Torner, L., Wise, F. \& Mihalache, D.
On multidimensional solitons and their legacy in contemporary atomic,
molecular and optical physics. \textit{J. Phys. B: At. Mol. Opt. Phys.}
\textbf{49}, 170502 (2016).}

\bibitem{spre} {\small Ablowitz, M. J. \& Musslimani, Z. H. Spectral
renormalization method for computing self-localized solutions to nonlinear
systems. \textit{Opt. Lett.} \textbf{30}, 2140-2142 (2005). }

\bibitem{stable1} {\small Kuznetsov, E. A., Rubenchik, A. M. \& Zakharov,
V. E. Soliton stability in plasmas and hydrodynamics. \textit{Phys. Rep}.
\textbf{142}, 103-165 (1986). }

\bibitem{stable2} {\small Yang, J. \textit{Nonlinear Waves in Integrable and
Nonintegrable Systems} (SIAM, Philadelphia, 2010). }


\end{thebibliography}
\end{document}